\documentclass{aa}
\usepackage[pdftex]{graphicx}
\usepackage{txfonts}
\usepackage{natbib}
\usepackage{longtable}
\usepackage[bookmarks,breaklinks]{hyperref}
   \hypersetup{
     colorlinks,
     citecolor=blue,
     filecolor=magenta,
     linkcolor=blue,
     urlcolor=cyan
}

\graphicspath{{Gamma/}}

\begin{document}

\title{$\gamma$-rays in flat-spectrum AGN: Revisiting the fast jet hypothesis with the CJF sample}

   \author{M. Karouzos\inst{1}
        \fnmsep\thanks{Member of the International Max Planck Research School (IMPRS) for Astronomy and Astrophysics at the Universities of Bonn and Cologne}
      \and S. Britzen\inst{1}
           \and A. Witzel\inst{1}
           \and J.A. Zensus\inst{1,2}
           \and A. Eckart\inst{2,1}
            }

   \institute{ Max-Planck-Institut f\"ur Radioastronomie, Auf dem H\"ugel 69, 53121 Bonn, Germany \\
            \email{mkarouzos@mpifr-bonn.mpg.de}
   \and
   I.Physikalisches Institut, Universit\"at zu K\"oln, Z\"ulpicher Str. 77, 50937 K\"oln, Germany
            }


   \date{Received / Accepted}

\abstract{The recent release of the First Fermi-LAT Source Catalog solidified the predominant association of extragalactic $\gamma$-ray emitters to active galaxies, in particular blazars. A tight connection between AGN jet kinematics and $\gamma$-ray properties has been argued for, attributing the energetic emission from active galaxies to their highly relativistic outflows.}
{We investigate the Caltech-Jodrell Bank flat-spectrum (CJF) sample to study the connection between AGN jet kinematics and their $\gamma$-ray properties. The high number of sources included in the sample, in addition to the excellent kinematic data available, allows us to investigate the origin of $\gamma$-ray emission in AGN.}
{We identify the CJF sources detected in $\gamma$-rays (by Fermi-LAT and EGRET). We use $\gamma$-ray luminosities and the available VLBI kinematic data to look for correlations between $\gamma$-ray and kinematic properties, as well as for differences between AGN classes (quasars, BL Lacs, radio galaxies). We also check the kinematics of the TeV sources in the CJF.}
{21.8\% of the CJF has been detected in the $\gamma$-ray regime. We find the detectability of BL Lacs significantly higher compared to quasars. $\gamma$-detected sources show a wider apparent jet velocity distribution compared to the non-detected ones, but the maxima of both distributions are at similar values. No strong link between $\gamma$-ray detectability and fast apparent jet speeds is found. A tentative correlation is found between $\gamma$-ray luminosity and maximum apparent jet speeds, stronger for BL Lac and $\gamma$-variable sources. We find non-radial jet motions to be important to $\gamma$-ray emission. We suggest two-zone, spine-sheath, models as a possible explanation to our results. We find 2 out of 4 CJF TeV sources show superluminal jet speeds, in contrast to previous studies.}
{}


   \maketitle

\section{Introduction}
\label{sec:intro}

Active galactic nuclei (AGN) are known for their emission over a broad range of the electromagnetic spectrum, reaching up to energy scales of $10^{12}$eV (e.g., \citealt{Punch1992}; \citealt{Neshpor1998}). This high energy end of the AGN spectrum is most probably created by inverse Compton scattering (either of the incident synchrotron photon field, self-scattering (SSC), of photons arising from the accretion disk, or of cosmic microwave background) from a population of relativistic electrons in the AGN jets (e.g., \citealt{Blandford1978}; \citealt{Ghisellini1985}; \citealt{Maraschi1992}; \citealt{Sikora1994}). Acceleration mechanisms, including acceleration from electric fields and acceleration in shocks, can boost the energy of the electron population enough so that $\gamma$-ray emission can be efficiently produced through inverse Compton scattering. The variability observed at these short wavelengths adds a further layer of complexity (e.g., \citealt{Hoyle1966}) and might require special geometrical and physical configurations to be explained (e.g., \citealt{Georganopoulos2005}; \citealt{Tavecchio2008}).

Most $\gamma$-ray radiation is shielded away from us by the Earth's atmosphere, except for very high energy (TeV) photons that produce atmospheric particle showers that can be detected and analyzed by Cherenkov telescopes. The launch of $\gamma$-ray satellites like the Compton $\gamma$-ray Observatory satellite (\citealt{Gehrels1993}; carrying the Energetic Gamma Ray Experiment Telescope, EGRET; \citealt{Kanbach1988}), the AGILE satellite (\citealt{Tavani2008}), and the Fermi $\gamma$-ray Space Telescope (\citealt{Atwood2009}) allows a detailed study of the $\gamma$-ray sky. In particular, the Fermi telescope offers the possibility of a continuous all-sky monitoring (using the Large Area Telescope, LAT) at much higher resolution and sensitivity than those of previous instruments. There are 1451 sources included in the first source catalog of the Fermi-LAT (\citealt{Abdo2010b}), 687 of which are identified/associated with active galaxies. Of these, 671 Fermi-LAT detected sources at high galactic latitudes ($|b|>10^\circ$) are included in the first catalog of active galactic nuclei (\citealt{Abdo2010}), associated with 709 AGN (multiple associations for some sources), comprising 300 BL Lac objects, 296 flat-spectrum radio quasars, 41 AGN of other type, and 72 of unknown class.

Of particular interest is the connection between the $\gamma$-ray and radio properties of AGN and the possibly common Doppler factor for both the radio and $\gamma$-ray emitting particles. The distribution of radio spectral indices of the Fermi-LAT detected AGN is consistent with a flat spectrum, although a minor tail of steeper spectrum sources exists (see Fig. 28 in \citealt{Abdo2010}), indicating the dominance of core-dominated sources as $\gamma$-ray emitters and reflecting the importance of the beaming effect for the production of this high-energy emission. The existence of a population of steeper spectrum sources, also detected by the Fermi-LAT, implies a more complex link between the $\gamma$ and radio properties of AGN.

\citet{Lister2009} study the kinematic properties of the 3-month Fermi-LAT detected sources of the MOJAVE sample. The authors find that $\gamma$-detected quasars show on average faster jets than their non-detected counterparts, in agreement with previous studies of EGRET-detected sources showing $\gamma$-detected sources to have preferentially higher Doppler beaming factors (e.g., \citealt{Jorstad2001b}; \citealt{Kellermann2004}). They also note that BL Lac sources show on average lower apparent speeds but are nevertheless preferentially detected by Fermi-LAT. They attribute this behavior to the spectral shape of BL Lacs and the possibility that BL Lacs have a higher intrinsic $\gamma$-ray to radio luminosity ratio than flat-spectrum QSOs, concluding that these results merit further investigation.

We use the Caltech-Jodrell Bank flat-spectrum (CJF; \citealt{Taylor1996}) sample of radio-loud AGN to test the putative connection between the $\gamma$-ray properties of AGN and their jet kinematics, in light of the recent release of the first Fermi-LAT AGN source catalog (\citealt{Abdo2010}) and in the context of the studies mentioned above. The paper is organized as follows: in Sect. \ref{sec:cjf} we describe the CJF sample, in Sect. \ref{sec:data} we present the available data, in Sect. \ref{sec:analysis} the data is analyzed and our results are presented, in Sect. \ref{sec:discussion} we discuss our results in the context of previous similar investigations, and we finally offer a summary and our conclusions in Sect. \ref{sec:conclusions}. Throughout the paper, we assume the cosmological parameters (from the first-year WMAP observations; \citealt{Spergel2003}) $H_{0}=71$ $km\:s^{-1}\:Mpc^{-1}$, $\Omega_{M}=0.27$, and $\Omega_{\Lambda}=0.73$.

\section{The CJF Sample}
\label{sec:cjf}

The CJF sample (\citealt{Taylor1996}) consists of 293 sources selected (see Table \ref{tab:cjfproperties}) from three different samples (for details see, \citealt{Britzen2007a}). The sources span a large redshift range (see Fig. 1 in \citealt{Britzen2008}), with the furthest object being at a redshift $z=3.889$ (1745+624; \citealt{Hook1995}) and the closest one at $z=0.0108$ (1146+596; \citealt{deVaucouleurs1991}). The average redshift of the sample is $z_{avg}=1.254$, with $z_{BL Lac,avg}=0.546$, $z_{RG,avg}=0.554$, and $z_{QSO,avg}=1.489$ for BL Lac objects, radio galaxies, and quasars, respectively. All the objects have been observed  with the VLBA and/or the global VLBI network. Each source has at least 3 epochs of observations and has been imaged and studied kinematically (\citealt{Britzen1999}; \citealt{Britzen2007a}; \citealt{Britzen2008}). The X-ray properties have also been studied and correlated with their VLBI properties (\citealt{Britzen2007b}). Finally, \citet{Karouzos2010} conduct a multi-wavelength study of the CJF in the context of the merger-driven evolution of galaxies.
\begin{center}
\begin{table}[h]
\caption{The CJF sample and its properties.}
\label{tab:cjfproperties}
\begin{tabular}{l c}
\hline\hline
\textbf{Frequency(MHz)}      &  4850 \\
\textbf{Flux lower limit @5GHz}    &  350mJy\\
\textbf{Spectral Index}      &  $\alpha_{1400}^{4850}\geq -0.5$ \\
\textbf{Declination}         &  $\delta\geq 35^{\circ}$  \\
\textbf{Galactic latitude}   &  $|b|\geq 10^{\circ}$ \\
\textbf{\# Quasars}           &  198 \\
\textbf{\# BL Lac }           &  32 \\
\textbf{\# Radio Galaxies}    &  52 \\
\textbf{\# Unclassified}      &  11  \\
\textbf{\# Total}             &  293 \\
\hline
\end{tabular}
\end{table}
\end{center}

The CJF offers an excellent tool to study the kinematics, given its large number of sources and the detailed investigation of the jet kinematics of all these sources. In light of the findings of \citet{Lister2009}, we are interested in investigating the putative connection between the $\gamma$-ray properties of the AGN in our sample and their jet kinematics, utilizing the extensive database of information already at our disposal (\citealt{Britzen2007a}; \citealt{Britzen2008}; Karouzos et al. 2010c, in prep.). We want to disentangle the different effects that might contribute in defining the $\gamma$-ray properties of an AGN and assess their relative importance. In particular, we want to test whether $\gamma$-detected AGN indeed show faster jets than their non-detected counterparts. We will also investigate the jet ridge line properties of the $\gamma$-detected sources, in relation to those of the non-detected ones (a detailed treatment of the CJF jet ridge lines will be presented in a separate paper).

\section{Data}
\label{sec:data}

The CJF sample is a statistically complete sample, under the selection criteria described in Sect. \ref{sec:cjf}. Radio data is therefore available for all the CJF sources, both single dish (at several frequencies) and interferometric. In addition, all CJF sources have been observed by ROSAT (see \citealt{Britzen2007a} for details). Finally, a substantial number of CJF sources has been recently detected in the $\gamma$-ray regime by the Fermi-LAT instrument. Below we describe the radio and $\gamma$-ray data available.

\subsection{Radio emission}

The CJF sample (Table \ref{tab:cjfproperties}) has been most extensively studied in the radio regime (e.g., \citealt{Taylor1996}; \citealt{Pearson1998}; \citealt{Britzen1999}; \citealt{Vermeulen2003}; \citealt{Pollack2003}; \citealt{Lowe2007}; \citealt{Britzen2007a}; \citealt{Britzen2007b}; \citealt{Britzen2008}). \citet{Britzen2008} develop a localized method for calculating the bending of the jet associated with individual components. The maximum of the distribution of local angles is at zero degrees, although a substantial fraction shows some bending ($0-40$ degrees). A few sources exhibit sharp bends of the order of $>50$ degrees (see Fig. 13 in \citealt{Britzen2008}).

\citet{Britzen2008} present an extensive analysis of the CJF jet kinematics, uniformly analyzing interferometric data for each CJF source, identifying individual components at each epoch, and studying their kinematic behavior. Although the CJF sample consists mostly of blazars, presumably highly beamed sources, the kinematical study of the sample shows a large number of sources having stationary, subluminal, or, at best, mildly superluminal outward velocities (e.g., see Fig. 15 in \citealt{Britzen2008}). Combined with a number of sources exhibiting inward moving components (e.g., 0600+422, 1751+441, 1543+517, \citealt{Britzen2007a}), these sources do not fit into the regular paradigm of outward, superluminaly moving components in blazar jets. Such peculiar kinematic behaviors can be readily explained by geometric effects, usually assuming a helical motion pattern combined with projection effects (e.g., \citealt{Zensus1995}; \citealt{Steffen1995}; \citealt{Steffen1995b}; \citealt{Lobanov2005}; \citealt{Roland2008}). Most of the CJF sources exhibit flux variability at different timescales, with several showing indications for quasi-periodicities in their radio lightcurves (see \citealt{Karouzos2010} for details).

\subsection{$\gamma$-ray emission}

EGRET detected 14 CJF sources and provided an upper limit for 50 more (\citealt{Fichtel1994}; \citealt{Hartman1999}). We note that, unlike for the radio and the X-rays, the $\gamma$-ray study of this sample is not complete since EGRET only did targeted observations of some of the CJF sources. Fermi-LAT (the renamed GLAST; \citealt{Atwood2009}) will provide a complete study of these sources through its all-sky survey. 61 CJF sources are included in the first catalog of AGN by the Fermi-LAT, after eleven months of observation (\citealt{Abdo2010}). Three additional sources are included in the third EGRET catalog (\citealt{Hartman1999}) but have not, as of yet, been detected by Fermi (0804+499, 2346+385, and 2351+456).

In total, we find 64 sources (21.8\% of the sample) being detected in the $\gamma$-ray regime, while for 40 (13.6\% of the sample) only an upper limit is reported from EGRET. Breaking down the number of detections, of the 64 $\gamma$-detected sources, 24 are classified as BL Lacs, 32 as quasars, and 5 radio galaxies. One $\gamma$-detected source remains unclassified. We mention here that there are 7 CJF sources detected by Fermi-LAT that have been classified differently by the Fermi-LAT team (see Table \ref{tab:gamma}). Three of these sources do not have kinematic data and therefore do not influence our result. For the remaining four, we retain the classification of \citet{Britzen2007b}. Six CJF sources belong to the group of few sources that have been detected in the TeV regime (Table \ref{tab:gammabend}).

Of the 61 CJF sources already detected by Fermi-LAT (Table \ref{tab:gamma}) we note that half of them are found to be $\gamma$-variable. In total, 31 sources (10.6\%) have been detected to be variable. We note for the following analysis we only use real detections, excluding upper limits information.

It should be noted that both the inclusion of a source in the LAT catalog, as well as a source being flagged as variable, depends on a number of factors, the combination of which make the catalog incomplete. The flux of a source (both in radio and $\gamma$-rays), its spectral shape, and its position on the sky (e.g., near the galactic plane) influence its possible association with an AGN. Moreover, as has already been noted in \citet{Abdo2010}, in order for a source to be classified as variable it needs to have sufficiently high $\gamma$-flux. Subsequent catalogs will surely improve upon the current associations and consequentially increase the number of $\gamma$-bright AGN.

\section{Analysis}
\label{sec:analysis}

\begin{figure}[h]
\begin{center}
  \includegraphics[width=0.4\textwidth,angle=0]{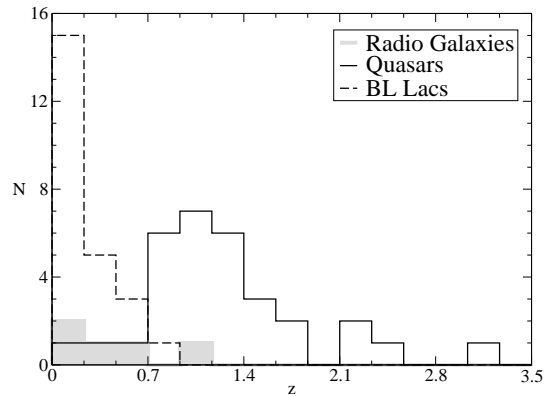}
  \caption[$\gamma$-detected CJF sources z distribution]{Redshift distribution of the Fermi-LAT detected (1st Catalog; \citealt{Abdo2010}) CJF sources. The three different classes of AGN are shown: BL Lacs (dashed empty line), quasars (continuous empty line), and radio galaxies (continuous filled line).}
  \label{fig:gamma_z_histo}
\end{center}
\end{figure}

As we described in the previous sections, we are interested in investigating a possible link between jet kinematics and $\gamma$-ray properties in the CJF sample. There are a couple of points that should be addressed before we do that. A first effect concerns the redshift distribution of our sources. Quasars show a redshift distribution peaking around z=1.3, while BL Lacs maximum is around z=0.2. In Fig. \ref{fig:gamma_z_histo} we show the redshift distribution of the CJF sources detected in $\gamma$-rays. The distribution follows the distribution of the sample, with the highest $\gamma$-detected source at redshift $z=3.044$. Secondly, the CJF sample contains predominantly quasars, with only a few percent of the sample being BL Lac objects and radio galaxies. This is a result of the selection criteria of the sample. 75\% of the CJF BL Lacs have been detected by Fermi-LAT, while only 17\% of the QSOs and 9.6\% of the RGs are included in the first year AGN catalog of the telescope (\citealt{Abdo2010}). Taking into account the number of QSOs at $z<1$ (56 sources), we get a $\gamma$-ray detectability of 16\% for the CJF quasars, as expected, significantly lower than BL Lacs.

\subsection{$\gamma$-ray luminosities}
\label{sec:analysis.gamma.lum}

$\gamma$-ray observations are based on photon counts per energy bin, unit surface, and unit time for each source. We are interested in translating the photon fluxes \textbf{(at the energy interval of 1-100 GeV)} to luminosities, in order to account for the distance dependence. We follow \citet{Thompson1996} to calculate $\gamma$-ray luminosities as follows: assuming an energy range $(E_1,E_2)$, then the integral photon flux is described as,
$$F(\Delta E)=\int_{E_{1}}^{E_{2}}AE^{-\alpha}dE,$$
where A is a normalization constant that can be expressed in terms of the integral flux F and $\alpha$ is the spectral index. For the Fermi-LAT instrument, a nominal value of 1-100 GeV is used for the energy range probed. Then, the energy flux is calculated from the following formula:
$$S(\Delta E)=\frac{1-\alpha}{2-\alpha}\cdot\frac{E_{2}^{2-\alpha}-E_{1}^{2-\alpha}}{E_{2}^{1-\alpha}-E_{1}^{1-\alpha}}F,$$
where F is the photon flux, the directly observed quantity (measured in photon counts per unit surface and time), and S is the energy flux (measured in ergs per unit surface and time). We use the median photon indices given in \citet{Abdo2010} for each source. It should be noted that, within this energy range, a break in the spectrum might exist. This would lead to an overestimation of the luminosity, especially for the QSO sub-sample (e.g., \citealt{Abdo2010c}).

Having calculated the sources luminosities, we can calculate average and median values for our sample (shown in Table \ref{tab:gamma_stats}. It should be noted that comparison of average and median values for non-Gaussian distributions is problematic. For the sake of continuation and uniformity with previous studies of the same topic we calculate both average and median values, although it should be seen as only indicative and interpreted in tandem with results from statistical tests like the K-S or Student's test.

Taking into account only QSOs at redshifts smaller than z=1, we can calculate the average and median values (assuming a normal distribution for the samples studied here, the uncertainty for the median values is calculated as 1.253 of the standard error of the mean) for this sub-sample ($47.09\pm0.11$ and $47.19\pm0.14\:\mathrm{erg}$, respectively; the values are given in the $\log {(\nu L_{\nu})}$ form). It can be seen that QSOs are consistently more luminous than the BL Lacs.

\begin{table}[h]
\begin{center}
\caption{Characteristic statistic values concerning the $\gamma$-ray luminosity (in logarithmic scale, measured in erg/s) of the CJF sources detected in the $\gamma$-ray regime by Fermi-LAT.}
\label{tab:gamma_stats}
\begin{tabular}{l| l| l l l|l l}
\multicolumn{7}{l}{}\\
\hline						
$\log {(\nu L_{\nu})_{\gamma}}$	&\textbf{All}	&\multicolumn{2}{c}{\textbf{QSO}} 	&\textbf{BL}	&\textbf{Var} &\textbf{Non-Var}	\\
 & &All &$z<1$ & &\multicolumn{2}{l}{}\\
\hline
\#	            &51	    &31	    &9      &14	       &30	    &21     \\
\textbf{Average}&47.26	&47.45	&46.09	&46.32     &47.41	&46.83  \\
\textbf{Error}	&0.07	&0.09	&0.11	&0.13      &0.09	&0.09   \\
\textbf{Median}	&46.63	&46.91	&47.19	&45.44     &46.73	&46.60  \\
\textbf{Error}	&0.09	&0.11	&0.14	&0.16      &0.11	&0.11  \\
\textbf{Max}	&48.58	&48.58	&47.50	&47.15     &48.58	&47.64  \\
\textbf{Min}	&43.67	&45.33	&45.33	&43.67     &43.89	&43.67  \\
\multicolumn{6}{l}{}\\
\hline
\end{tabular}
\tablefoot{We use the flux between 1 and 100 GeV to calculate the $\gamma$-ray luminosity. We give average and values with uncertainties, maximum, and minimum values for the whole sub-sample, QSOs, QSOs at $z<1$, BL Lacs, variable, and non-variable sources. Given the small number of RGs (5 sources) detected, we do not calculate separate statistics for that sub-sample.}
\end{center}
\end{table}

The luminosity of a source at longer wavelengths might play a role in deciding its $\gamma$-ray properties. To that end, we compare the luminosities of the $\gamma$-detected and non-detected QSOs at $z<1$, at 5 and 30 GHz (single dish), at optical (V band), and in the soft X-rays data from \citealt{Britzen2007a}, \citealt{Taylor1996}, and references therein). The $\gamma$-detected sources are consistently more luminous in the radio and optical regime, but are fainter in the X-rays. As was described previously, up to soft X-rays the emission is thought to be produced by the synchrotron mechanism and therefore $\gamma$-detected sources show a stronger synchrotron component than their non-detected counterparts. This might in turn be linked to the putative inverse Compton process usually employed to explain the production of $\gamma$-ray emission in AGN jets. Conversely, $\gamma$-ray detection of a source implies that the inverse Compton hump of its SED is shifted towards higher energies, compared to non-detected sources. It is therefore to be expected that $\gamma$-detected sources are actually weaker in the soft X-rays than their non-detected counterparts. A larger sample of $\gamma$-detected sources, along with a closely matched (in terms of luminosity and redshift) control sample is required to test this scenario.

\begin{figure}[h]
\begin{center}
  \includegraphics[width=0.5\textwidth,angle=0]{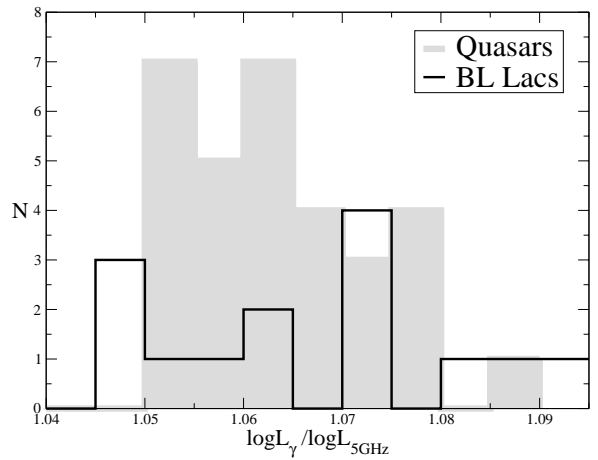}
  \caption[$\gamma$-to-radio ratio: QSOs vs. BL Lacs]{Histogram comparing the distributions of $\gamma$-to-radio luminosity ratios for BL Lacs and QSOs. We use $\gamma$-ray luminosities derived from the Fermi-LAT photon fluxes given in the first Fermi-LAT source catalog (\citealt{Abdo2010b}) and radio luminosities at 5 GHz derived from single dish energy fluxes from \citet{Taylor1996}.}
  \label{fig:gamma_spectral_histo}
\end{center}
\end{figure}

Concerning the relative importance of the synchrotron and inverse Compton components, we calculate the $\gamma$-to-radio luminosity ratio for the $\gamma$-detected CJF sources (see Fig. \ref{fig:gamma_spectral_histo}).

In  Fig. \ref{fig:gamma_spectral_histo} we show the distribution of this ratio for quasars and BL Lacs (including both variable and non-variable sources). It can be seen that BL Lacs show marginally higher $\gamma$-to-radio ratios than quasars, with the BL Lac distribution peaking around 1.075 compared to the quasar one peaking around 1.055. The fact that we include both variable and non-variable sources might influence our results. Given the observational bias that sources classified as $\gamma$-variable are on average brighter than the non-variable ones and that there are more variable BL Lacs than non-variable ones, the effect observed in Fig. \ref{fig:gamma_spectral_histo} might in part be due to the same observational bias discussed previously. A two sample K-S test for BL Lacs and quasars is inconclusive at to whether their $\gamma$-to-radio radio distributions are drawn from different parent distributions. An obvious caveat of this comparison is that $\gamma$ and radio observations are not contemporaneous.

\subsection{Apparent VLBI Jet Component Velocities and $\gamma$-ray emission}

\begin{figure*}[htp]
\begin{center}
  \includegraphics[width=0.8\textwidth,angle=0]{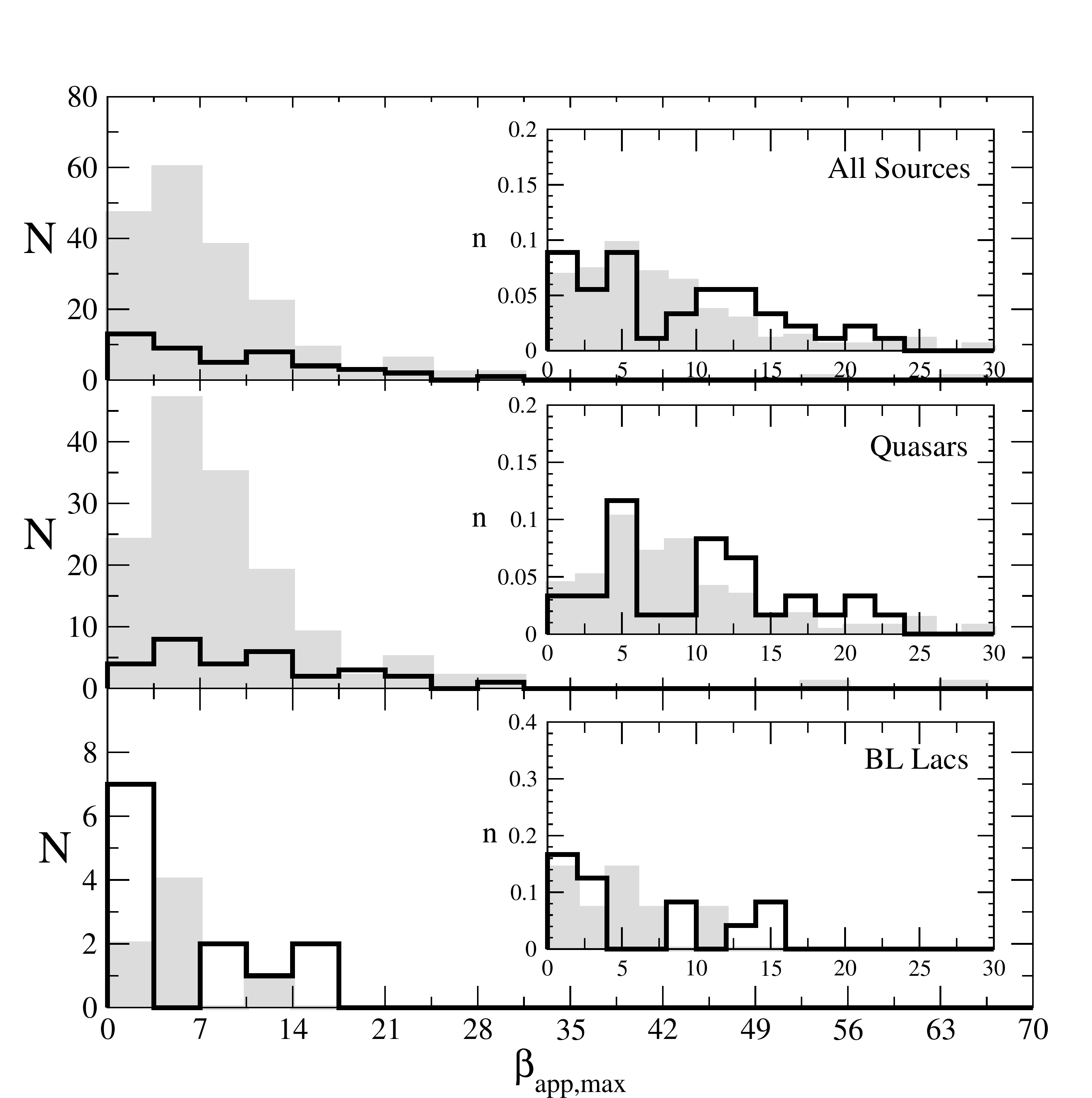}
  \caption[$\beta_{app,max}$ distribution: $\gamma$-detected and non-detected sources]{The distribution of the maximum apparent velocities $\beta_{app,max}$ for sources that have been detected in the $\gamma$-ray regime (black line) and for those that have not (grey blocks). We show the distributions for all sources (upper panel), quasars (middle panel), and BL Lacs (lower panel). The insets show the normalized to surface area unity distributions for each case and for apparent speeds up to 30c. We use the kinematic data from \citet{Britzen2008}.}
  \label{fig:doppler_gamma}
\end{center}
\end{figure*}

We characterize the kinematics of the CJF sources by the maximum observed component velocity, $\beta_{app,max}$, of each source. This maximum component velocity \textbf{depends on} both the orientation of the source (viewing angle) and the intrinsic properties of the jet itself. We investigate whether the distribution of maximum apparent component speeds differs between $\gamma$-detected and non-detected sources.  For all CJF sources with available redshift, \citet{Britzen2008} have calculated the apparent total component velocities for all the identified components in their VLBI jets. We identify the component with the maximum apparent speed in each source and we plot the distribution of these maximum apparent speeds in Fig. \ref{fig:doppler_gamma}, for $\gamma$-detected sources (black line) and non-detected ones (grey blocks).

In the upper panel of Fig. \ref{fig:doppler_gamma}, we plot the distribution of the maximum apparent velocities for the 191 CJF sources that have not been detected at $\gamma$-ray wavelengths (grey blocks) and the 45 sources that have (black line). The maximum of the distribution for the non-detections is found in the [3.5,7] bin. For the detections the maximum lies in the [0,3.5] bin. As can be seen in the inset of the upper panel (inset plots are normalized to surface area one, as we are interested in the relative distributions of the two sub-samples), assuming smaller bins, the maximum of the $\gamma$-detected sources breaks down to two maxima in the [0,2] and [4,6] bins. The non-detected sources distribution is peaked in the [4,6] bin. $\gamma$-detected sources have a more extended distribution, towards higher velocities, showing a possible secondary peak in the [10,12] and [12,14] bins. A K-S test between $\gamma$-detected and non-detected sources $\beta_{app,max}$ distributions does not give a conclusive answer (93.7\% confidence that the two sub-samples are different).

In the middle panel of Fig. \ref{fig:doppler_gamma} we show the same plots as before but only for those sources classified as quasars (same notation as before). The distribution here is markedly different from before. Both populations ($\gamma$-detected and non-detected) have their distribution shifted towards higher velocities, both of them peaking in the [4,6] bin. The $\gamma$-detected QSOs show a possible secondary peak in the [10,12] bin (secondary to primary ratio of 0.71) and again appear to show a larger fraction of the total number of sources at higher velocities. Non-detected quasars have a possible secondary peak in the [8,10] bin (secondary to primary ratio of 0.8). The highest velocities for quasars are found in sources that have not been detected at $\gamma$-ray wavelengths. A K-S test gives a low probability (of 95.4\%) that the two samples are significantly different. Similarly, in the lower panel of Fig. \ref{fig:doppler_gamma} we show the distributions (absolute and normalized) for BL Lac objects. Both $\gamma$-detected and non-detected BL Lacs have their distribution maxima in the [0,2] bin, with $\gamma$-detected BL Lacs having a considerably more extended distribution, reaching to higher velocities, compared to their non-detected counterparts. Compared to quasars, the BL Lac velocity distribution is shifted towards lower values. A K-S test for the distribution of $\beta_{app,max}$ for $\gamma$-detected quasars and BL Lacs gives a 97.7\% confidence that they are drawn from different parent samples. It should be noted however that the number of $\gamma$-detected BL Lacs with available redshift and kinematic information is small and therefore our analysis for BL Lacs is probably affected by low number statistics.

We also calculate average and median values of the maximum apparent velocity for $\gamma$-ray detected CJF sources and for those not detected ($9.0\pm0.8$ and $8.8\pm1.0$, compared to $8.1\pm0.4$ and $6.4\pm0.5$, respectively). A Student's t-test is inconclusive (49\% for the null hypothesis). A further note concerns the redshift distribution of the two sub-samples. The $\gamma$-detected sources have a lower average redshift (0.967) compared to their non-detected counterparts (1.175). It is known that there appears to be a dependence between apparent velocity and redshift (as noted by \citealt{Cohen2007}; \citealt{Britzen2008}; \citealt{Lister2009b}; Karouzos et al. 2010c, in prep.). This effect reinforces the difference in average and median values seen between $\gamma$-detected and non-detected sources and implies that $\gamma$-detected sources indeed have higher maximum apparent velocities.

Finally, for the $\gamma$-ray detected sources, we distinguish between quasars and BL Lac objects. We find that quasars exhibit considerably higher average and median maximum apparent velocities ($10.8\pm1.0$ and $10.2\pm1.2$, respectively, compared to $6.1\pm1.5$ and $2.9\pm1.9$). Redshift effects might again be influencing our results. We select those QSOs at redshifts lower than 1 and calculate $\beta_{app,max}$ average and median values. We find that the sub-sample of local QSOs shows, within the statistical errors, the same average values as the BL Lacs ($6.9\pm1.2$) but considerably higher median value ($6.1\pm1.5$).

\begin{figure}[h]
\begin{center}
  \includegraphics[width=0.4\textwidth,angle=0]{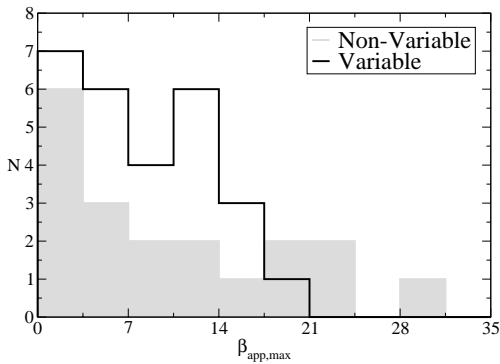}
  \caption[$\beta_{app,max}$ distribution: Variable and non-variable $\gamma$-detected sources]{Distribution of the maximum apparent velocities $\beta_{app,max}$ for $\gamma$-detected sources that are variable (black line) and for those that are not (grey blocks). We use the kinematic data from \citet{Britzen2008}.}
  \label{fig:gamma_beta_var_histo}
\end{center}
\end{figure}

In Fig. \ref{fig:gamma_beta_var_histo} we compare the $\beta_{app,max}$ distribution of variable (27 sources; average redshift $z_{avg}=0.94\pm0.09$) and non-variable (19 sources; average redshift $z_{avg}=0.94\pm0.10$) $\gamma$-detected CJF sources. In the following we refer to $\gamma$-ray variability, not taking into account possible variability in other wavelength regimes. Non-variable $\gamma$-detected sources show a more extended distribution than the variable ones, reaching the highest velocities ($\sim$30c). Both distributions show their main maximum in the [0,3.5] bin. Variable sources show a possible secondary maximum in the [10.5,14] bin (with a secondary to primary ratio of 0.86). A K-S test for the two distributions does not provide a conclusive result.

As in the previous cases, we again calculate the statistical properties of the two sub-samples. Variable $\gamma$-detected sources show similar $\beta_{app,max}$ average value to their non-variable counterparts, within the statistical errors ($7.9\pm0.8$ and $8.9\pm1.7$, respectively). When checking the median values however, the variable sub-sample shows substantially higher value than the non-variable one ($8.6\pm1.0$, compared to $4.6\pm2.1$). Both sub-samples show similar redshift distributions, therefore we do not expect any redshift effect influencing our result.

\begin{figure}[htp]
\begin{center}
  \includegraphics[width=0.5\textwidth,angle=0]{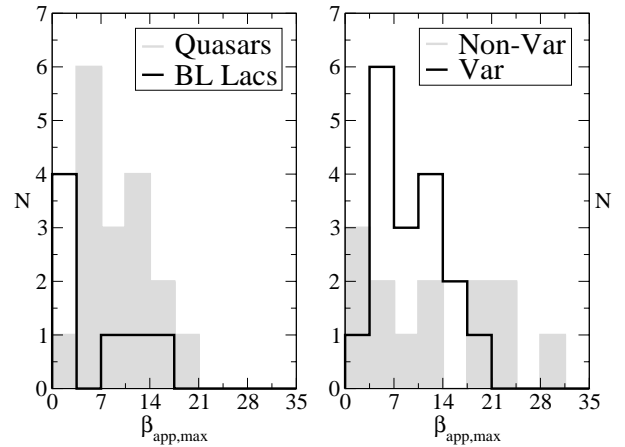}
  \caption[$\beta_{app,max}$ distribution: Variable vs. Non-Variable / QSOs vs. BL Lacs]{ Left panel: distributions of the maximum apparent velocities $\beta_{app,max}$ for $\gamma$-variable BL Lacs (black line) quasars (grey blocks). Right panel: $\gamma$-variable (black line) and non-variable (grey blocks) quasars. We use the kinematic data from \citet{Britzen2008}.}
  \label{fig:gamma_beta_var_BL_QSO_histo}
\end{center}
\end{figure}

In Fig. \ref{fig:gamma_beta_var_BL_QSO_histo} we compare the $\beta_{app,max}$ distributions of $\gamma$-detected, variable quasars and $\gamma$-detected, variable BL Lacs (left panel) and $\gamma$-detected variable quasars and $\gamma$-detected non-variable quasars (right panel). These two sub-samples (variable sources and quasar sources) are the largest of the different $\gamma$-detected sub-samples (e.g., non-variable sources, BL Lac sources, etc.) and therefore are chosen to check differences between quasars and BL Lacs, and variability, in an isolated manner.

We find that quasars (left panel of Fig. \ref{fig:gamma_beta_var_BL_QSO_histo}) show the maximum of their distribution at higher values than BL Lacs ([3.5,7] bin compared to [0,3.5]). Given the small number for each group of sources, a K-S test is not applied. For variable and non-variable sources the comparison is not so straightforward (right panel of Fig. \ref{fig:gamma_beta_var_BL_QSO_histo}). Non-variable $\gamma$-detected quasars show a main maximum in the [0,3.5] bin, lower than their variable counterparts. However, they also show a more extended distribution reaching higher $\beta_{app,max}$ values than their variable counterparts. A K-S test gives inconclusive results (23\% probability that the sub-samples are the same). Small number statistics, as well as the biased detection of bright sources as variable, affect our results and therefore we can not give a definitive answer concerning the relative importance of the two effects. There is evidence that $\gamma$-ray emitting, variable quasars show statistically higher apparent velocities.

The above analysis implies that there are two effects, possible dependent, that correlate with the jet kinematic properties of $\gamma$-emitting AGN, (1) $\gamma$-variability and (2) the classification of the source (BL Lac or quasar). The latter can be associated either to the difference in un-beamed luminosity between BL Lacs and quasars, or possibly difference in viewing angles (e.g., \citealt{Lahteenmaki1999}; \citealt{Hovatta2009}). For the number of sources available here, we can not robustly decouple these two effects. Given the available data, we do not find any significant connection between $\gamma$-ray detection and fast moving jet components, as has been argued by other authors. We rather see that the $\beta_{app,max}$ distribution is more strongly dependent on the type of object, i.e., BL Lac or quasar classification (see Sect. \ref{sec:discussion} for a discussion on this).

\subsection{$\gamma$-ray Luminosities and Apparent Jet Component Velocities}
\begin{figure*}[htp]
\begin{center}
  \includegraphics[width=0.8\textwidth,angle=0]{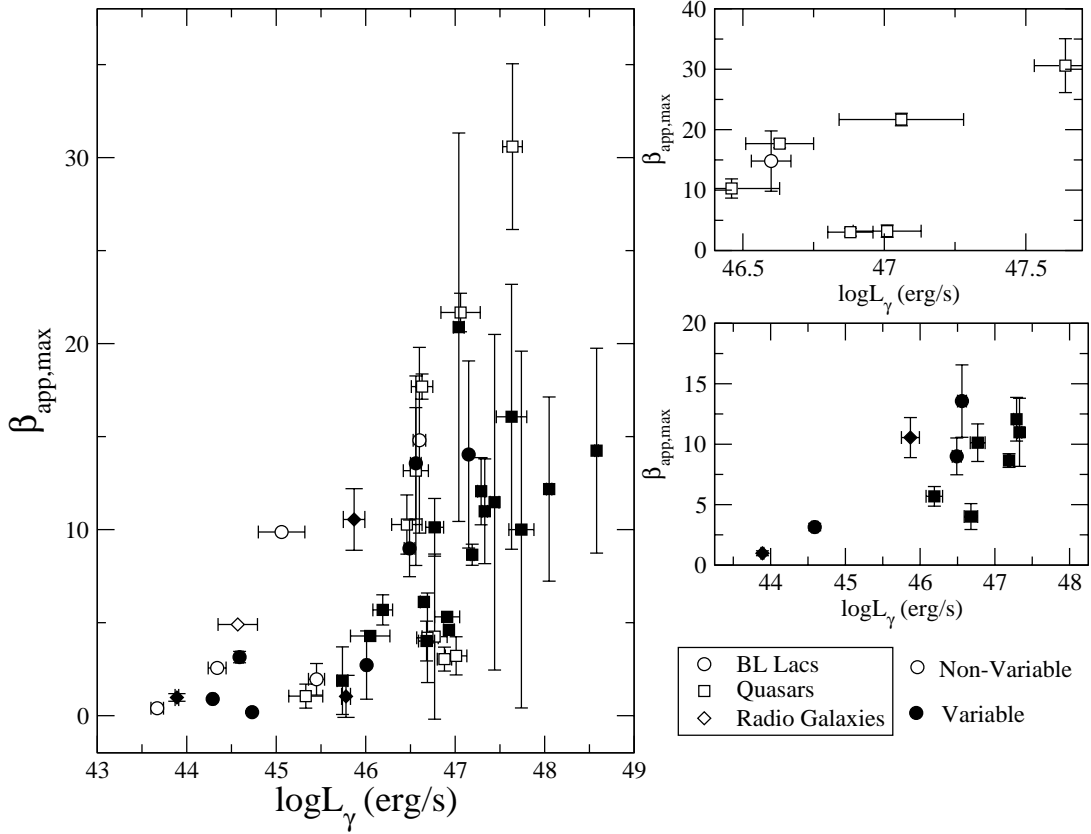}
  \caption[$\beta_{app,max}$ and $\gamma$-ray luminosity]{Maximum apparent component velocities for all $\gamma$-ray detected CJF sources as a function of their $\gamma$-ray luminosity, for all sources (left panel) and for high significance values of $\beta_{app,max}$ (right panels; see text for details). We differentiate between $\gamma$-variable (filled symbols) and non-variable sources (open symbols), and between the different AGN classes, i.e., BL Lacs (circles), quasars (squares), and radio galaxies (diamonds). We use the kinematic data from \citet{Britzen2008}.}
  \label{fig:gamma_beta_lumin}
\end{center}
\end{figure*}

In the previous section we probed the putative connection between the apparent brightness of $\gamma$-rays and the jet kinematics -i.e., the beaming mechanism-, as reflected in the differences between the $\beta_{app,max}$ distribution of quasars and BL Lacs. Another way to approach this is by looking for a possible direct correlation between the apparent velocities measured in the AGN jets and their $\gamma$-ray luminosity. This is shown in Fig. \ref{fig:gamma_beta_lumin}.

As can be seen in Fig. \ref{fig:gamma_beta_lumin}, there appears to exist a correlation between the $\beta_{app,max}$ of a source and its $\gamma$-ray luminosity. Variable sources appear to cluster closer to the implied trend, with non-variable QSOs deviating the most. We find Spearman correlation coefficients of 0.68 and 0.77 (both at a significance of $>99.999\%$), respectively. Given the mutual dependence of luminosity and jet component apparent velocity to the redshift of a source, we also calculate the Pearson product moment partial correlation coefficients r($L_\gamma$ $\beta_{app,max}$,z)\footnote{For this we use the Web tool: Wessa P., (2008), Partial Correlation (v1.0.4) in Free Statistics Software (v1.1.23-r6), Office for Research Development and Education, URL http://www.wessa.net/rwasp\_partialcorrelation.wasp/.}. While for the whole sample we get a relatively low partial correlation coefficient (0.43 at a significance of 99.8\%), the correlation for variable sources persists (0.68 at a significance of $>99.9\%$). Given the expected degree of scatter in the data, the correlation coefficients combined with the calculated significance imply that the trend seen in Fig. \ref{fig:gamma_beta_lumin} is indeed true.
We note that some of the apparent velocities show relatively large errors, therefore, in the right panels of Fig. \ref{fig:gamma_beta_lumin}, we plot only sources for which $\beta_{app,max}\ge 3\sigma$. We also separate variable sources from non-variable ones. It becomes clear that the non-variable sources show the most scatter. For the variable sources, we calculate a Spearman correlation coefficient of 0.69 (at a 99.1\% significance level), lower than the coefficient we got when fitting all the sources. Calculating the partial correlation coefficient for the same sub-sample gives a smaller correlation coefficient (0.61 at a 98.1\% significance). We also investigate the same correlation for individual classes of objects. The strongest correlations are seen for both $\gamma$-variable BL Lacs (partial correlation coefficient of 0.68 at 95.1\% significance) and quasars (partial correlation coefficient of 0.69 at 99.9\% significance). The differences seen between the different classes of AGN, as well as between variable and non-variable sources reveal a complicated picture. We shall discuss the robustness and implications of our results more extensively in Sect. \ref{sec:discussion}.

\subsection{$\gamma$-ray and jet ridge line properties}
\begin{figure}[htp]
\begin{center}
  \includegraphics[width=0.4\textwidth,angle=0]{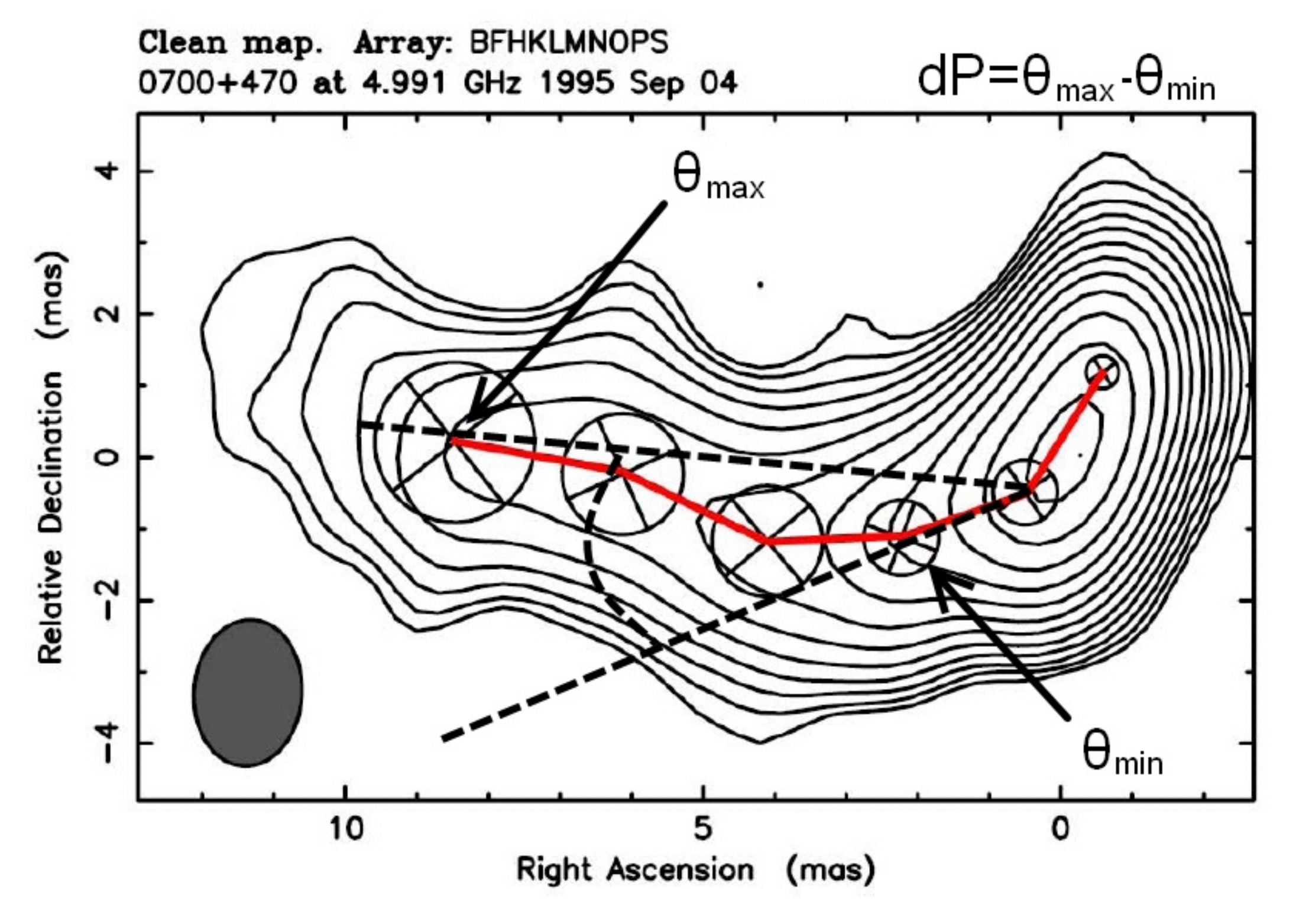}
  \caption[Jet Ridge Line Width Example]{Example of the definition of the jet width for the source 0700+470. With arrows we show the components with the minimum and maximum position angles at that epoch. The continuous line represents the jet ridge line of 0700+470 (see the text for a definition) at the same epoch. Map from \citet{Britzen2007a}.}
  \label{fig:ridge_line_width_example}
\end{center}
\end{figure}

The currently accepted paradigm for the jet kinematics of flat-spectrum sources (i.e., core-dominated AGN) includes superluminaly outward moving components, usually interpreted in the context of a specific projected geometry combined with relativistic effects, due to the intrinsically high speeds of the bulk flow. Recent detailed kinematic studies of the parsec-scale jets of BL Lac objects (e.g., 1803+784, \citealt{Britzen2010}; 0716+714, \citealt{Britzen2009}; 0735+178, \citealt{Britzen2010b}) have however revealed a rather different kinematic scheme for their jets: BL Lac jet components are predominantly stationary with respect to the core but change their position angle significantly, essentially reflecting an important transverse component in their movement. In addition, their jet ridge lines, defined as the line that linearly connects the projected positions of all components at a certain epoch, show significant temporal evolution, at times forming very wide flow funnels. In an accompanying paper, we shall present a statistical investigation of the jet ridge lines of the CJF sources (Karouzos et al. 2010c, in prep.).
We are interested in investigating a possible correlation between the jet ridge line properties of a source and its appearance at $\gamma$-ray wavelengths. We shall therefore briefly outline the method used to analyze the CJF jet ridge lines. We can define three measures that probe both the radial and transverse motion of the jet ridge line of a source, (1) the jet ridge line width, dP, (2) the jet ridge line width evolution, $\Delta P$, and (3) the linear evolution of the jet ridge line, $\Delta\ell$. The dP is defined as the position angle difference between the components with the maximum and minimum position angle at a given epoch, measured in degrees (see Fig. \ref{fig:ridge_line_width_example} for an example). The $\Delta P$ is derived between two successive epochs as the difference between the jet ridge line widths at these epochs (measured in degrees per unit time). The $\Delta\ell$ is derived as the sum of the linear displacements, along the vector of their motion, of all jet components, across all available epochs (measured in parsecs per unit time and per component) and resembles an average component apparent speed across all epochs.

\begin{figure}[htp]
\begin{center}
  \includegraphics[width=0.5\textwidth,angle=0]{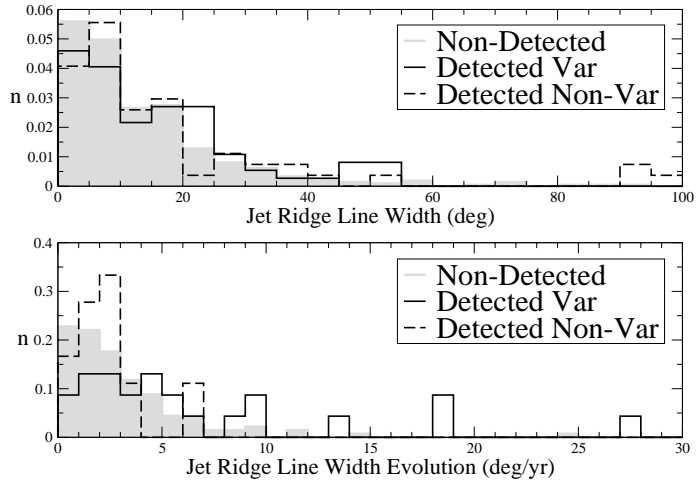}
  \caption[Jet ridge line properties (dP and $\Delta P$) and $\gamma$-rays]{Histograms of jet ridge line width, dP, and width evolution, $\Delta P$, for $\gamma$-detected variables sources (continuous black line), $\gamma$-detected non-variable sources (dashed black line), and $\gamma$-ray non-detected sources (grey blocks). Given the large difference in absolute numbers, we plot histograms normalized to surface area unity. We use data from Karouzos et al. (2010c, in prep.).}
  \label{fig:gamma_width}
\end{center}
\end{figure}

We compare these three measures of the jet ridge line properties for $\gamma$-detected and non-detected sources. We find that $\gamma$-detected sources show significantly wider jet ridge lines, both in average and median values ($16.9\pm1.0^\circ$ and $11.1\pm1.2^\circ$, respectively), compared to the non-detected ones ($13.6\pm0.4^\circ$ and $9.3\pm0.5^\circ$). A Student's t-test gives a 97.6\% probability that the two average values are significantly different. The difference between the two distributions can be also clearly seen in the upper panel of Fig. \ref{fig:gamma_width}: the maximum of the jet ridge line width distribution of the $\gamma$-detected (non-variable) sources is shifted to higher values (in the [5,10] bin), compared to their non-detected counterparts (which have their maximum in the [0,5] bin).

Furthermore, $\gamma$-detected sources are found to show stronger evolution of their widths, both in average and median values ($4.7\pm0.5$ deg/yr and $2.5\pm0.6$ deg/yr, respectively), compared to their non-detected counterparts ($3.13\pm0.18$ deg/yr and $2.26\pm0.22$ deg/yr). A Student's t-test confirms at a 98.1\% confidence that the two averages are significantly different. In the lower panel of Fig. \ref{fig:gamma_width} we show the normalized distribution of the width evolution values for $\gamma$-detected and non-detected sources. The $\gamma$-detected distribution is peaked in the [2,3] bin, compared to the non-detected one, peaking in the [0,1] bin.

We can also distinguish between variable and non-variable $\gamma$-detected CJF sources and investigate their jet ridge line width evolution in this context. Variable sources are found to have significantly stronger evolving jet ridge line widths, at a 4$\sigma$ significance level, both in average and median values ($6.7\pm1.0$ and $4.4\pm1.2$ deg/yr, compared to $2.4\pm0.3$ and $2.2\pm0.4$ deg/yr). A Student's t-test confirms this result (at a 98.8\% confidence). In Fig. \ref{fig:gamma_width} (lower panel) we plot the distribution of $\gamma$-detected variable and non-variable sources. The non-variable sources distribution is fairly confined to lower values, whereas variable sources extend up to the highest values of width evolution. This implies that viewing angle changes at parsec scales of AGN jets are linked to variability in the $\gamma$-ray regime, reflecting a possibly evolving jet at even smaller scales (connected to the timescales associated with $\gamma$-ray variability).

We finally compare how the jet ridge line evolves, in linear terms, in $\gamma$-detected and non-detected sources, essentially this time looking for a possible link between $\gamma$-ray brightness of a source and its average jet component apparent speed, instead of the maximum apparent speed. As we already discussed previously, the redshift distributions of the two sub-samples are fairly similar. We do not expect therefore any redshift-induced effects, related to linear distances, to influence our results. We find that both in average and median values, the two sub-samples show similar values of linear evolution ($0.47\pm0.03$ and $0.40\pm0.04$ pc/yr/comp, respectively, for the $\gamma$-detected sources, compared to $0.440\pm0.020$ and $0.377\pm0.025$ pc/yr/comp for the non-detected ones).

The same behavior is seen when comparing between $\gamma$-detected variable and non-variable sources. It should be noted however, that the median value for $\gamma$-detected variable sources is fairly higher than the one for non-variable sources ($0.43\pm0.04$ pc/yr/comp compared to $0.291\pm0.025$ pc/yr/comp, respectively). This discrepancy between average and median values implies a large scattering within our data that might be posing some limitations to the robustness of our results. Aside from that, we confirm our finding from before, i.e., that the link between fast apparent jet speeds and the $\gamma$-ray detectability of a source is questionable. It appears that, although there are certainly indications that higher speed sources are preferentially $\gamma$-ray emitters, some other effect might play a more important role in defining the $\gamma$-ray properties of a source. Our results concerning the width and width evolution comparison between $\gamma$-detected and non-detected sources imply that non-radial motion -i.e., motion transverse to the jet symmetry axis- is important with regard to $\gamma$-ray production. This is probably tightly connected to the viewing angle effect, also presumably reflected in the kinematic differences between different AGN types.

\subsection{TeV sources}
28  extragalactic sources have been identified to be emitters in the TeV regime (for an updated list of these sources and references, see http://tevcat.uchicago.edu/). Six CJF sources (5 BL Lac objects, 1 radio galaxy) are observed and detected in TeV. All are variable in $\gamma$-rays and optical, five in the infrared, while five of them are variable in radio and four in the X-rays.

Extremely high energetic photons coming from AGN (TeV sources, see Table \ref{tab:gamma}) can be a result of several different mechanisms, some of them taking place in the AGN jets (for a discussion on high energy $\gamma$-ray emission from AGN see e.g., \citealt{Montigny1995}). We check the kinematics of the TeV sources in the CJF (Table \ref{tab:gammabend}; data from \citealt{Britzen2008}). We give the misalignment angle between the parsec and the kiloparsec scale (as calculated by \citealt{Britzen2007b}). We also list the bending of the parsec scale jet (as calculated by \citealt{Britzen2008}) for all identified components. Finally we give the maximum apparent total jet component speed, $\beta_{app,max}$. We find no apparent correlation between the bending of the jet and the $\gamma$-ray flux. 2 out of 5 sources exhibit superluminal motion in their VLBI jets. Previous studies of TeV emitting high-frequency peaked BL Lacs exhibit slow jet speeds (e.g., \citealt{Piner2008}; \citealt{Britzen2009}). 0219+428 and 2200+420 do not belong to that category of objects.

\begin{table*}
\begin{center}
\caption{Jet morphology and kinematics of TeV detected CJF sources.}
\label{tab:gammabend}
\begin{tabular}{l l l l l l l}
\hline\hline
\textbf{Source}	&	\textbf{Type}	&\textbf{z}	&$\mathbf{\Delta PA}$	&\textbf{Bend} &\textbf{$\beta_{tot,max}$}		 &$\mathbf{\gamma}$\textbf{-rays	}\\
	            &		            &		    & (deg)		            & (deg) &		        &($10^{-8} ph\:cm^{-2}\:s^{-1}$)	     \\
	\hline
0219+428&	BL 	&0.444		&8		&24	 &$14\pm5$	&$2.49\pm0.10$	 \\
	    &		&		    &		&4   &     & 			 \\
	    &		&		    &		&1   &   	&		 \\
	\hline
0316+413&   G   &0.018      &20     &52  &$0.98\pm0.20$ &$1.73\pm0.08$  \\
        &       &           &       &85  &              &   \\
        &       &           &       &83     &   &   \\
        &       &           &       &24 &   &   \\
    \hline
0716+714&	BL 	&		    &75		&3	 &	&$1.31\pm0.07$	  \\
	    &		&		    &		&7   &   &      			  \\
	    &		&		    &		&3   &  &                 \\
	\hline
1101+384&	BL 	&0.031		&13		&2	 &$0.19\pm0.07$	&$2.61\pm0.10$ \\
\hline
1652+398&	BL 	&0.034    	&83		&10	 &0.90	&$0.83\pm0.06$		  \\
	    &		&		    &		&40  &  &                 \\
	    &		&		    &		&10  &  &                 \\
	    &		&		    &		&3   & &                 \\
	\hline	
2200+420&	BL 	&0.069		&30		&25	 &$3.2\pm0.3$	&$0.71\pm0.06$		  \\
	    &		&		    &		& 16 &                 \\
	    &		&		    &		& 47 &                 \\
\hline
\end{tabular}
\tablefoot{Columns (1)-(3) give the IAU name, the type, and the redshift of the source, Col. (4) gives the misalignment angle between the pc and kpc-scale jet (from \citealt{Britzen2007b}), Col. (5) gives the bending of the identified components of the pc-scale jet (for details see \citealt{Britzen2008}), Col. (6) gives the maximum apparent total speed for the VLBI jet of the source, and Col. (7) gives the source's $\gamma$-ray flux between 1 and 100 GeV (Tabel \ref{tab:gamma}).}
\end{center}
\end{table*}

\section{Discussion}
\label{sec:discussion}

It is interesting to compare our results with similar studies using different samples of AGN. As we already mentioned in Sect. \ref{sec:intro}, a number of authors have claimed a close connection between $\gamma$-ray bright AGN and large $\beta_{app,max}$ (e.g., \citealt{Jorstad2001b}; \citealt{Kellermann2004}; \citealt{Lister2009}). \citet{Lister2009}, in particular, examine the kinematic properties of the MOJAVE sources (\citealt{Lister2009b}) in light of the, then recently, published 3-month bright AGN Fermi-LAT list. In short, they find that (1) the $\gamma$-detected MOJAVE quasars have the peak of their $\beta_{app,max}$ distribution around 10-15c, quite a higher value compared to the non-detected ones ($\sim$5c). They also find that (2) $\gamma$-variable sources show higher apparent speeds. Finally, although BL Lacs in their sample have lower redshifts and slow median jet speeds (6c), (3) they are preferentially detected by Fermi-LAT. Similarly to \citet{Lister2009}, \citet{Jorstad2001b} and \citet{Kellermann2004} also conclude that $\gamma$-detected AGN have significantly higher apparent speeds than their non-detected counterparts.

The results presented here are somewhat different. Although we do find a possible secondary maximum around 10-15c for the $\gamma$-detected sources, the primary maximum of both distributions appears to be around 5c, with $\gamma$-detected sources actually showing a double maximum in the [0,2] and [4,6] bins. The $\beta_{app,max}$ distribution of the sources is more strongly dependent on the classification of the source as a quasar or a BL Lac, and whether the source appears variable, rather than it being detected at $\gamma$-ray wavelengths. Given the larger number of sources in the CJF, unlike \citet{Lister2009b}, we can treat the different AGN population separately. In particular, as we showed in the previous section, quasars show higher apparent jet speeds than BL Lacs, regardless of their $\gamma$-ray properties. A common feature for $\gamma$-detected sources is that their $\beta_{app,max}$ distribution appears more extended, with larger percentage of the sources at higher speeds, compared to those of non-detected sources. Coming to point (2) -i.e., that $\gamma$-variable sources show higher apparent jet speeds- our results agree, as we showed for the case of variable and non-variable $\gamma$-detected quasars. However this does not come as a surprise in light of our recent findings from the statistical analysis of the CJF jet ridge lines. In short, we find that variable sources (throughout the electromagnetic spectrum) show stronger linear evolution of their jet ridge lines, a measure essentially reflecting their apparent jet speeds (Karouzos et al. 2010, in prep.). Therefore, the higher apparent speeds of $\gamma$-detected variable sources can probably be linked to the variability, rather than to the $\gamma$-ray brightness itself.

Finally, concerning point (3), we also find that BL Lac objects are preferentially detected by Fermi-LAT, even after correcting for the different redshift distribution of quasars and BL Lacs. \citet{Lister2009} argue that BL Lacs might be preferentially detected by Fermi-LAT because of their flatter spectrum, i.e., their higher $\gamma$-to-radio ratio, compared to quasars. As was shown in Sect. \ref{sec:analysis.gamma.lum}, this is indeed true.

A further point of interest, that has not so far been investigated, concerns the apparent correlation seen between the $\gamma$-ray luminosity of a source and its maximum apparent jet speed. As we showed in Fig. \ref{fig:gamma_beta_lumin} such a correlation exists, with variable sources falling closer to the implied trend. We need to contemplate whether the inclusion of more sources with potentially lower $\gamma$-ray luminosities would destroy this correlation. Looking at the CJF sources not yet detected by Fermi-LAT, 35 of them show significant ($>3\sigma$) apparent speeds above 10c, while only 10 show speeds larger than 20c. Of these 10 sources only 3 are classified as high quality component fits (see \citealt{Britzen2008} for details). Together with the high median redshift of the high-speed $\gamma$ non-detected CJF sources, the above suggest that the correlation seen in Fig. \ref{fig:gamma_beta_lumin} is robust.

Although our results are generally in agreement with previous studies, there are some differences that merit further interpretation. Of particular interest is the fact that we do not find a strong connection between fast moving components and $\gamma$-detected sources, as opposed to previous works. As we already discussed in Sect. \ref{sec:intro}, one of the models proposed for the origin of $\gamma$-ray emission in AGN involves a spine-sheath geometry, where a high velocity spine is embedded in a slower moving sheath. According to these models, the ultra-relativistic spine gives rise to the $\gamma$-ray emission that we observe. Assuming that the jet as a whole becomes gradually more opaque at lower observing frequencies, we could then argue that, given the lower frequency that the CJF sample has been observed in, we are only probing the (transversely) outer regions of the jet and therefore we only recover the slower motions linked to the sheath of the jet. Conversely, the MOJAVE sample (15 GHz) as well as the samples used by both \citet{Jorstad2001b} (42, 22, and occasionally 15 and 8.4 GHz) and \citet{Kellermann2004} (15 GHz) are all at higher observing frequencies and thus more sensitive to the inner layers of the jet. It is therefore plausible that these studies have recovered the faster speeds linked to the ultra-relativistic spine and the production of the observed $\gamma$-ray emission. An additional effect related to the higher frequency observation of the MOJAVE programme is the distances from the core probed. At 15 GHz, the higher resolution of the VLBI observations allow the investigation of smaller spatial scales, closer to the core, and therefore possibly more closely connected to $\gamma$-ray production processes. Finally, given the limited number of epochs available for some of the CJF sources, combined with the low observing frequency, it can be argued that the fastest moving components might have been missed or blended together with slower components in the CJF kinematic study, therefore resulting to, on average, slower component speeds for the whole sample. Detailed studies of individual sources at both higher temporal and spatial resolutions (e.g., 1803+784, \citealt{Britzen2010}; 0716+714, \citealt{Britzen2009}; etc.) have shown that this is not the case (if not showing the presence of the reverse effect).

An alternative scenario would be that indeed $\gamma$-ray emission, although boosted in high-Doppler jets, is not necessarily coupled to fast jet components speeds. Given that \citet{Lister2009} use the three month bright source list for their study, it is possible that only the $\gamma$-brightest, hence the most strongly beamed, sources are detected. Therefore there is a strong bias towards sources with higher Doppler factor jets. Including sources from the first, 11-month, Fermi-LAT catalog allows us to probe lower-flux sources (given the higher sensitivity achieved after 11-months of observations) and therefore the bias towards the most highly beamed sources is lifted. This scenario however is unlikely since (1) by using only the sources included in the 3-month bright sources Fermi-LAT list and our CJF apparent speed data, we still recover a low-velocity component in the $\beta_{app,max}$ distribution of the $\gamma$-detected sources and (2) \citet{Savolainen2010}, after calculating Doppler and Lorentz factors of Fermi-LAT detected AGN (3-month list), find a notable absence of sources at the smallest (and therefore most highly beamed) co-moving viewing angles. Distinguishing between either scenarios will require a larger sample of $\gamma$-detected AGN with uniform, multi-frequency kinematic data available.

A last possibility that might be affecting the kinematic properties of the $\gamma$-detected sources pertains to the relativistic effect of sources observed at a viewing angle inside the $1/\Gamma$ cone, where $\Gamma$ is the Lorentz factor of the flow, having lower apparent speeds. \citet{Hovatta2009} calculate median values for Lorentz factors and viewing angles for different AGN samples. The authors find for BL Lac objects a median Lorentz factor of $\Gamma=10.29$ and a median viewing angle of $\theta=5.24$, while for flat-spectrum quasars they find 16.24 and 3.37, respectively.  These values suggest that both sub-classes of AGN are found inside the $1/\Gamma$ cone. Given a certain combination of intrinsic speeds and viewing angles, sources could occur that are beamed enough to be picked up by Fermi-LAT but at the same time exhibit slower apparent jet speeds. The fact that the same population of sources is not seen in the MOJAVE sample makes this explanation debatable.

A further layer of complexity is introduced when we consider the $\gamma$-ray variability of most of the $\gamma$-detected sources. If the $\gamma$-ray emission from AGN is of transient nature, related to a relativistic shock propagating down the pc-scale jet, this implies that many sources are missed from such an investigation, simply because they are observed during their quiescent state. In addition to this, the non-contemporaneity of the kinematic and $\gamma$-ray data introduces further noise to plots like Figs. \ref{fig:doppler_gamma} and \ref{fig:gamma_beta_lumin}. At present we can not sidestep these two caveats.

The last point that needs to be discussed concerns the connection between the jet ridge line width to the $\gamma$-ray properties of a source. \citet{Pushkarev2009} study the jet opening angles of the MOJAVE sample again in the context of the 3-month bright source list of the Fermi-LAT. The authors find that $\gamma$-bright sources have larger apparent opening angles, but in their co-moving frame both $\gamma$-bright and faint sources show similar opening angle distributions. This is in turn interpreted as evidence that $\gamma$-detected sources are seen at smaller viewing angles than the non-detected ones. Our results agree with these findings, as we find that $\gamma$-detected sources exhibit wider jet ridge lines compared to their non-detected counterparts. It should however be noted that preferential detection of BL Lacs at $\gamma$-ray wavelengths combined with our results from the statistical analysis of the CJF jet ridge lines (i.e., that BL Lacs have apparently wider jet ridge lines; Karouzos et al. 2010c, in prep.) might introduce a spurious effect in such an analysis. It is also interesting to note that, according to our findings from Sect. \ref{sec:analysis}, $\gamma$-detected sources also exhibit large changes in their jet ridge line widths, larger than the non-detected ones. That would introduce a noise factor when comparing the width distributions of $\gamma$-detected and non-detected sources. This in turn implies that $\gamma$-detected sources might in reality be even wider than inferred by the distribution shown in Fig. \ref{fig:gamma_width}.

It is interesting to contemplate on how a two-zone model might accommodate for the apparently important non-radial component of the component trajectories, as implied by the significantly larger jet ridge line width evolution shown by $\gamma$-detected sources. A helical structure of AGN jets combined with the smaller viewing angles expected for $\gamma$-bright sources might offer a plausible explanation. Such a geometry has been used to explain non-ballistic trajectories for a number of AGN jets (e.g., 3C345, \citealt{Steffen1995}; 1803+784, \citealt{Steffen1995b}; 1633+382, \citealt{Liu2010}). One could speculate that a helically structured sheath, enclosing an ultra-relativistic (straight) spine flow, could give rise to the kinematic behavior observed in the CJF sources, while explaining the $\gamma$-ray properties of these sources. Given the still small number of available CJF sources detected by Fermi-LAT, it is, for the time being, not possible to decouple and assess the individual importance of these effects.

\section{Conclusions}
\label{sec:conclusions}

In the previous sections we have investigated the connection of the $\gamma$-ray properties of the CJF sources to the morphologic and kinematic properties of their jets. In summary we find:
\begin{itemize}
 \item 21.8\% of the CJF sample is detected at $\gamma$-ray wavelengths (either from EGRET or the Fermi-LAT; three EGRET associations not included in the 11 month catalog)
    \item BL Lacs appear to be preferentially detected in the $\gamma$-ray regime. Taking into account the difference of the redshift distributions of QSOs and BL Lacs in our sample, we still get fairly different detection ratios between the two classes (16\% and 75\%, respectively).
    \item after calculating the $\gamma$-ray luminosities of both QSOs and BL Lacs, and taking into account the redshift effects of our flux-limited sample, we find that QSOs appear more luminous at $\gamma$-ray wavelengths than BL Lacs.
    \item $\gamma$-detected sources (regardless of classification) show the peak of their $\beta_{app,max}$ distribution at similar values as their non-detected counterparts but show a more extended distribution towards higher jet component speeds. A K-S test results in an inconclusive answer, giving a 93.7\% confidence that the two sub-samples are significantly different.
    \item when considering QSOs and BL Lacs separately, we still find that $\gamma$-detected and non-detected sources show fairly similar distributions, with $\gamma$-detected sources showing more extended distributions towards higher values.
    \item comparing between $\gamma$-detected QSOs and BL Lacs, after accounting for redshift effects, both sub-samples show on average and within statistical errors the same apparent speeds.
     \item we find a tentative correlation between $\beta_{app,max}$ and $\gamma$-ray luminosity. The correlation is stronger for BL Lac objects and for $\gamma$-variable sources, with non-variable QSOs deviating the most from the implied trend.
    \item we find that $\gamma$-detected sources have significantly wider jet ridge lines than their non-detected counterparts. We also find that $\gamma$-detected sources show stronger jet ridge line width evolution than non-detected ones.
    \item we find no significant difference in terms of linear evolution of the jet ridge lines between $\gamma$-detected and non-detected sources.
    \item we find no direct link between highly bent jets and TeV emission. Furthermore, we note that 2 out of 4 TeV CJF sources with $\beta_{app,max}$ information show superluminal speeds, unlike previous studies.
 \end{itemize}

From our analysis it becomes clear that there is a number of factors influencing whether a source is luminous in the $\gamma$-ray regime or not. Although it is tempting to think in terms of $\gamma$-loud and quiet objects, the picture is surely more complicated. As has been recently demonstrated for the radio divide (radio-loud and radio-quiet objects) there is a number of sources that actually seem to populate an intermediate space, implying a rather continuous distribution. It is possible that with more sensitive $\gamma$-ray telescopes a greater number of $\gamma$-ray emitting sources will be recovered. Of course the question still remains as to what it is that differentiates the $\gamma$-quiet (or -faint) sources from those already detected by the Fermi-LAT and EGRET missions. Our analysis, combined with previous studies on this question, indicates that the viewing angle and $\gamma$-variability are what make some sources to be $\gamma$-ray luminous and others not. However, as has been shown by \citet{Savolainen2010}, the picture may be more complicated than that. Given the relatively weak link between jet apparent speeds and $\gamma$-detected sources that we find in this work, as well as the different picture arising at higher observing frequencies, a spine-sheath configuration scenario could offer a plausible explanation, where the most energetic emission - coming from an ultra-relativistic flow- originates in the spine of the AGN outflow and is therefore partly or fully obscured at lower observing frequencies. Finally, the link between the width and width evolution of the jet ridge lines of our sources to their $\gamma$-ray properties implies that transverse, non-radial, motions in the jet might be important in this context, probably also related to the $\gamma$-variability detected for almost half of the $\gamma$-detected AGN in our sample. It is obvious that a larger sample of $\gamma$-detected AGN with complete, multi-frequency kinematic data will allow us to investigate the above effects in a more robust manner. Furthermore, the investigation of both individual objects and statistical samples (like the CJF and the MOJAVE, e.g., \citealt{Chang2010}; Chang et al. 2010, in prep.) with detailed modeling of their SEDs will surely shed light to the actual processes producing the $\gamma$-ray emission.

\begin{acknowledgements}
M. Karouzos was supported for this research through a stipend from the International Max Planck Research School (IMPRS) for Astronomy and Astrophysics. The authors thank the anonymous referee for helpful suggestions that have improved this paper. M.K. wants to also thank Tuomas Savolainen and Mar Mezcua for insightful discussions and comments that significantly improved this manuscript. This research has made use of the NASA/IPAC Extragalactic Database (NED) which is operated by the Jet Propulsion Laboratory, California Institute of Technology, under contract with the National Aeronautics and Space Administration. This research has made use of NASA's Astrophysics Data System Bibliographic Services.
\end{acknowledgements}

\bibliographystyle{aa}
\bibliography{bibtex}

\begin{thebibliography}{53}
\expandafter\ifx\csname natexlab\endcsname\relax\def\natexlab#1{#1}\fi

\bibitem[{{Abdo} {et~al.}(2010{\natexlab{a}}){Abdo}, {Ackermann}, {Ajello},
  {Allafort}, {Antolini}, {Atwood}, {Axelsson}, {Baldini}, {Ballet},
  {Barbiellini}, {Bastieri}, {Baughman}, {Bechtol}, {Bellazzini}, {Berenji},
  {Blandford}, {Bloom}, {Bogart}, {Bonamente}, {Borgland}, {Bouvier},
  {Bregeon}, {Brez}, {Brigida}, {Bruel}, {Buehler}, {Burnett}, {Buson},
  {Caliandro}, {Cameron}, {Cannon}, {Caraveo}, {Carrigan}, {Casandjian},
  {Cavazzuti}, {Cecchi}, {{\c C}elik}, {Celotti}, {Charles}, {Chekhtman},
  {Chen}, {Cheung}, {Chiang}, {Ciprini}, {Claus}, {Cohen-Tanugi}, {Conrad},
  {Costamante}, {Cotter}, {Cutini}, {D'Elia}, {Dermer}, {de Angelis}, {de
  Palma}, {De Rosa}, {Digel}, {Silva}, {Drell}, {Dubois}, {Dumora}, {Escande},
  {Farnier}, {Favuzzi}, {Fegan}, {Ferrara}, {Focke}, {Fortin}, {Frailis},
  {Fukazawa}, {Funk}, {Fusco}, {Gargano}, {Gasparrini}, {Gehrels}, {Germani},
  {Giebels}, {Giglietto}, {Giommi}, {Giordano}, {Giroletti}, {Glanzman},
  {Godfrey}, {Grandi}, {Grenier}, {Grondin}, {Grove}, {Guiriec}, {Hadasch},
  {Harding}, {Hayashida}, {Hays}, {Healey}, {Hill}, {Horan}, {Hughes},
  {Iafrate}, {Itoh}, {J{\'o}hannesson}, {Johnson}, {Johnson}, {Johnson},
  {Johnson}, {Kamae}, {Katagiri}, {Kataoka}, {Kawai}, {Kerr}, {Kn{\"o}dlseder},
  {Kuss}, {Lande}, {Latronico}, {Lavalley}, {Lemoine-Goumard}, {Llena Garde},
  {Longo}, {Loparco}, {Lott}, {Lovellette}, {Lubrano}, {Madejski}, {Makeev},
  {Malaguti}, {Massaro}, {Mazziotta}, {McConville}, {McEnery}, {McGlynn},
  {Michelson}, {Mitthumsiri}, {Mizuno}, {Moiseev}, {Monte}, {Monzani},
  {Morselli}, {Moskalenko}, {Murgia}, {Nolan}, {Norris}, {Nuss}, {Ohno},
  {Ohsugi}, {Omodei}, {Orlando}, {Ormes}, {Ozaki}, {Paneque}, {Panetta},
  {Parent}, {Pelassa}, {Pepe}, {Pesce-Rollins}, {Piranomonte}, {Piron},
  {Porter}, {Rain{\`o}}, {Rando}, {Razzano}, {Reimer}, {Reimer}, {Reposeur},
  {Ripken}, {Ritz}, {Rodriguez}, {Romani}, {Roth}, {Ryde}, {Sadrozinski},
  {Sanchez}, {Sander}, {Saz Parkinson}, {Scargle}, {Sgr{\`o}}, {Shaw},
  {Siskind}, {Smith}, {Spandre}, {Spinelli}, {Starck}, {Stawarz}, {Strickman},
  {Suson}, {Tajima}, {Takahashi}, {Takahashi}, {Tanaka}, {Taylor}, {Thayer},
  {Thayer}, {Thompson}, {Tibaldo}, {Torres}, {Tosti}, {Tramacere}, {Ubertini},
  {Uchiyama}, {Usher}, {Vasileiou}, {Vilchez}, {Villata}, {Vitale}, {Waite},
  {Wallace}, {Wang}, {Winer}, {Wood}, {Yang}, {Ylinen}, \&
  {Ziegler}}]{Abdo2010b}
{Abdo}, A.~A., {Ackermann}, M., {Ajello}, M., {et~al.} 2010{\natexlab{a}},
  \apj, 715, 429

\bibitem[{{Abdo} {et~al.}(2010{\natexlab{b}}){Abdo}, {Ackermann}, {Ajello},
  {Allafort}, {Antolini}, {Atwood}, {Axelsson}, {Baldini}, {Ballet},
  {Barbiellini}, {Bastieri}, {Baughman}, {Bechtol}, {Bellazzini}, {Berenji},
  {Blandford}, {Bloom}, {Bogart}, {Bonamente}, {Borgland}, {Bouvier},
  {Bregeon}, {Brez}, {Brigida}, {Bruel}, {Buehler}, {Burnett}, {Buson},
  {Caliandro}, {Cameron}, {Cannon}, {Caraveo}, {Carrigan}, {Casandjian},
  {Cavazzuti}, {Cecchi}, {{\c C}elik}, {Celotti}, {Charles}, {Chekhtman},
  {Chen}, {Cheung}, {Chiang}, {Ciprini}, {Claus}, {Cohen-Tanugi}, {Conrad},
  {Costamante}, {Cotter}, {Cutini}, {D'Elia}, {Dermer}, {de Angelis}, {de
  Palma}, {De Rosa}, {Digel}, {Silva}, {Drell}, {Dubois}, {Dumora}, {Escande},
  {Farnier}, {Favuzzi}, {Fegan}, {Ferrara}, {Focke}, {Fortin}, {Frailis},
  {Fukazawa}, {Funk}, {Fusco}, {Gargano}, {Gasparrini}, {Gehrels}, {Germani},
  {Giebels}, {Giglietto}, {Giommi}, {Giordano}, {Giroletti}, {Glanzman},
  {Godfrey}, {Grandi}, {Grenier}, {Grondin}, {Grove}, {Guiriec}, {Hadasch},
  {Harding}, {Hayashida}, {Hays}, {Healey}, {Hill}, {Horan}, {Hughes},
  {Iafrate}, {Itoh}, {J{\'o}hannesson}, {Johnson}, {Johnson}, {Johnson},
  {Johnson}, {Kamae}, {Katagiri}, {Kataoka}, {Kawai}, {Kerr}, {Kn{\"o}dlseder},
  {Kuss}, {Lande}, {Latronico}, {Lavalley}, {Lemoine-Goumard}, {Llena Garde},
  {Longo}, {Loparco}, {Lott}, {Lovellette}, {Lubrano}, {Madejski}, {Makeev},
  {Malaguti}, {Massaro}, {Mazziotta}, {McConville}, {McEnery}, {McGlynn},
  {Michelson}, {Mitthumsiri}, {Mizuno}, {Moiseev}, {Monte}, {Monzani},
  {Morselli}, {Moskalenko}, {Murgia}, {Nolan}, {Norris}, {Nuss}, {Ohno},
  {Ohsugi}, {Omodei}, {Orlando}, {Ormes}, {Ozaki}, {Paneque}, {Panetta},
  {Parent}, {Pelassa}, {Pepe}, {Pesce-Rollins}, {Piranomonte}, {Piron},
  {Porter}, {Rain{\`o}}, {Rando}, {Razzano}, {Reimer}, {Reimer}, {Reposeur},
  {Ripken}, {Ritz}, {Rodriguez}, {Romani}, {Roth}, {Ryde}, {Sadrozinski},
  {Sanchez}, {Sander}, {Saz Parkinson}, {Scargle}, {Sgr{\`o}}, {Shaw},
  {Siskind}, {Smith}, {Spandre}, {Spinelli}, {Starck}, {Stawarz}, {Strickman},
  {Suson}, {Tajima}, {Takahashi}, {Takahashi}, {Tanaka}, {Taylor}, {Thayer},
  {Thayer}, {Thompson}, {Tibaldo}, {Torres}, {Tosti}, {Tramacere}, {Ubertini},
  {Uchiyama}, {Usher}, {Vasileiou}, {Vilchez}, {Villata}, {Vitale}, {Waite},
  {Wallace}, {Wang}, {Winer}, {Wood}, {Yang}, {Ylinen}, \&
  {Ziegler}}]{Abdo2010}
{Abdo}, A.~A., {Ackermann}, M., {Ajello}, M., {et~al.} 2010{\natexlab{b}},
  \apj, 715, 429

\bibitem[{{Abdo} {et~al.}(2010{\natexlab{c}}){Abdo}, {Ackermann}, {Ajello},
  {Atwood}, {Axelsson}, {Baldini}, {Ballet}, {Barbiellini}, {Bastieri},
  {Bechtol}, {Bellazzini}, {Berenji}, {Blandford}, {Bloom}, {Bonamente},
  {Borgland}, {Bouvier}, {Bregeon}, {Brez}, {Brigida}, {Bruel}, {Burnett},
  {Buson}, {Caliandro}, {Cameron}, {Caraveo}, {Carrigan}, {Casandjian},
  {Cavazzuti}, {Cecchi}, {{\c C}elik}, {Charles}, {Chekhtman}, {Cheung},
  {Chiang}, {Ciprini}, {Claus}, {Cohen-Tanugi}, {Conrad}, {Cutini}, {Dermer},
  {de Angelis}, {de Palma}, {Digel}, {Silva}, {Drell}, {Dubois}, {Dumora},
  {Farnier}, {Favuzzi}, {Fegan}, {Focke}, {Fortin}, {Frailis}, {Fukazawa},
  {Funk}, {Fusco}, {Gargano}, {Gasparrini}, {Gehrels}, {Germani}, {Giebels},
  {Giglietto}, {Giommi}, {Giordano}, {Glanzman}, {Godfrey}, {Grenier},
  {Grondin}, {Grove}, {Guillemot}, {Guiriec}, {Harding}, {Hartman},
  {Hayashida}, {Hays}, {Healey}, {Horan}, {Hughes}, {Jackson},
  {J{\'o}hannesson}, {Johnson}, {Johnson}, {Kamae}, {Katagiri}, {Kataoka},
  {Kawai}, {Kerr}, {Kn{\"o}dlseder}, {Kuss}, {Lande}, {Latronico},
  {Lemoine-Goumard}, {Longo}, {Loparco}, {Lott}, {Lovellette}, {Lubrano},
  {Madejski}, {Makeev}, {Mazziotta}, {McConville}, {McEnery}, {Meurer},
  {Michelson}, {Mitthumsiri}, {Mizuno}, {Moiseev}, {Monte}, {Monzani},
  {Morselli}, {Moskalenko}, {Murgia}, {Nolan}, {Norris}, {Nuss}, {Ohsugi},
  {Omodei}, {Orlando}, {Ormes}, {Paneque}, {Panetta}, {Parent}, {Pelassa},
  {Pepe}, {Persic}, {Pesce-Rollins}, {Piron}, {Porter}, {Rain{\`o}}, {Rando},
  {Razzano}, {Reimer}, {Reimer}, {Reposeur}, {Ritz}, {Rochester}, {Rodriguez},
  {Romani}, {Roth}, {Ryde}, {Sadrozinski}, {Sanchez}, {Sander}, {Saz
  Parkinson}, {Scargle}, {Sgr{\`o}}, {Siskind}, {Smith}, {Smith}, {Spandre},
  {Spinelli}, {Strickman}, {Suson}, {Tajima}, {Takahashi}, {Takahashi},
  {Tanaka}, {Thayer}, {Thayer}, {Thompson}, {Tibaldo}, {Torres}, {Tosti},
  {Tramacere}, {Uchiyama}, {Usher}, {Vasileiou}, {Vilchez}, {Villata},
  {Vitale}, {Waite}, {Wang}, {Winer}, {Wood}, {Ylinen}, \&
  {Ziegler}}]{Abdo2010c}
{Abdo}, A.~A., {Ackermann}, M., {Ajello}, M., {et~al.} 2010{\natexlab{c}},
  \apj, 710, 1271

\bibitem[{{Atwood} {et~al.}(2009){Atwood}, {Abdo}, {Ackermann}, {Althouse},
  {Anderson}, {Axelsson}, {Baldini}, {Ballet}, {Band}, {Barbiellini},
  {Bartelt}, {Bastieri}, {Baughman}, {Bechtol}, {B{\'e}d{\'e}r{\`e}de},
  {Bellardi}, {Bellazzini}, {Berenji}, {Bignami}, {Bisello}, {Bissaldi},
  {Blandford}, {Bloom}, {Bogart}, {Bonamente}, {Bonnell}, {Borgland},
  {Bouvier}, {Bregeon}, {Brez}, {Brigida}, {Bruel}, {Burnett}, {Busetto},
  {Caliandro}, {Cameron}, {Caraveo}, {Carius}, {Carlson}, {Casandjian},
  {Cavazzuti}, {Ceccanti}, {Cecchi}, {Charles}, {Chekhtman}, {Cheung},
  {Chiang}, {Chipaux}, {Cillis}, {Ciprini}, {Claus}, {Cohen-Tanugi},
  {Condamoor}, {Conrad}, {Corbet}, {Corucci}, {Costamante}, {Cutini}, {Davis},
  {Decotigny}, {DeKlotz}, {Dermer}, {de Angelis}, {Digel}, {do Couto e Silva},
  {Drell}, {Dubois}, {Dumora}, {Edmonds}, {Fabiani}, {Farnier}, {Favuzzi},
  {Flath}, {Fleury}, {Focke}, {Funk}, {Fusco}, {Gargano}, {Gasparrini},
  {Gehrels}, {Gentit}, {Germani}, {Giebels}, {Giglietto}, {Giommi}, {Giordano},
  {Glanzman}, {Godfrey}, {Grenier}, {Grondin}, {Grove}, {Guillemot}, {Guiriec},
  {Haller}, {Harding}, {Hart}, {Hays}, {Healey}, {Hirayama}, {Hjalmarsdotter},
  {Horn}, {Hughes}, {J{\'o}hannesson}, {Johansson}, {Johnson}, {Johnson},
  {Johnson}, {Johnson}, {Kamae}, {Katagiri}, {Kataoka}, {Kavelaars}, {Kawai},
  {Kelly}, {Kerr}, {Klamra}, {Kn{\"o}dlseder}, {Kocian}, {Komin}, {Kuehn},
  {Kuss}, {Landriu}, {Latronico}, {Lee}, {Lee}, {Lemoine-Goumard}, {Lionetto},
  {Longo}, {Loparco}, {Lott}, {Lovellette}, {Lubrano}, {Madejski}, {Makeev},
  {Marangelli}, {Massai}, {Mazziotta}, {McEnery}, {Menon}, {Meurer},
  {Michelson}, {Minuti}, {Mirizzi}, {Mitthumsiri}, {Mizuno}, {Moiseev},
  {Monte}, {Monzani}, {Moretti}, {Morselli}, {Moskalenko}, {Murgia},
  {Nakamori}, {Nishino}, {Nolan}, {Norris}, {Nuss}, {Ohno}, {Ohsugi}, {Omodei},
  {Orlando}, {Ormes}, {Paccagnella}, {Paneque}, {Panetta}, {Parent}, {Pearce},
  {Pepe}, {Perazzo}, {Pesce-Rollins}, {Picozza}, {Pieri}, {Pinchera}, {Piron},
  {Porter}, {Poupard}, {Rain{\`o}}, {Rando}, {Rapposelli}, {Razzano}, {Reimer},
  {Reimer}, {Reposeur}, {Reyes}, {Ritz}, {Rochester}, {Rodriguez}, {Romani},
  {Roth}, {Russell}, {Ryde}, {Sabatini}, {Sadrozinski}, {Sanchez}, {Sander},
  {Sapozhnikov}, {Parkinson}, {Scargle}, {Schalk}, {Scolieri}, {Sgr{\`o}},
  {Share}, {Shaw}, {Shimokawabe}, {Shrader}, {Sierpowska-Bartosik}, {Siskind},
  {Smith}, {Smith}, {Spandre}, {Spinelli}, {Starck}, {Stephens}, {Strickman},
  {Strong}, {Suson}, {Tajima}, {Takahashi}, {Takahashi}, {Tanaka}, {Tenze},
  {Tether}, {Thayer}, {Thayer}, {Thompson}, {Tibaldo}, {Tibolla}, {Torres},
  {Tosti}, {Tramacere}, {Turri}, {Usher}, {Vilchez}, {Vitale}, {Wang},
  {Watters}, {Winer}, {Wood}, {Ylinen}, \& {Ziegler}}]{Atwood2009}
{Atwood}, W.~B., {Abdo}, A.~A., {Ackermann}, M., {et~al.} 2009, \apj, 697, 1071

\bibitem[{{Blandford} \& {Icke}(1978)}]{Blandford1978}
{Blandford}, R.~D. \& {Icke}, V. 1978, \mnras, 185, 527

\bibitem[{{Britzen} {et~al.}(2007{\natexlab{a}}){Britzen}, {Brinkmann},
  {Campbell}, {Gliozzi}, {Readhead}, {Browne}, \& {Wilkinson}}]{Britzen2007b}
{Britzen}, S., {Brinkmann}, W., {Campbell}, R.~M., {et~al.} 2007{\natexlab{a}},
  \aap, 476, 759

\bibitem[{{Britzen} {et~al.}(2009){Britzen}, {Kam}, {Witzel}, {Agudo}, {Aller},
  {Aller}, {Karouzos}, {Eckart}, \& {Zensus}}]{Britzen2009}
{Britzen}, S., {Kam}, V.~A., {Witzel}, A., {et~al.} 2009, \aap, 508, 1205

\bibitem[{{Britzen} {et~al.}(2010{\natexlab{a}}){Britzen}, {Kudryavtseva},
  {Witzel}, {Campbell}, {Ros}, {Karouzos}, {Mehta}, {Aller}, {Aller},
  {Beckert}, \& {Zensus}}]{Britzen2010}
{Britzen}, S., {Kudryavtseva}, N.~A., {Witzel}, A., {et~al.}
  2010{\natexlab{a}}, \aap, 511, A57+

\bibitem[{{Britzen} {et~al.}(2008){Britzen}, {Vermeulen}, {Campbell}, {Taylor},
  {Pearson}, {Readhead}, {Xu}, {Browne}, {Henstock}, \&
  {Wilkinson}}]{Britzen2008}
{Britzen}, S., {Vermeulen}, R.~C., {Campbell}, R.~M., {et~al.} 2008, \aap, 484,
  119

\bibitem[{{Britzen} {et~al.}(2007{\natexlab{b}}){Britzen}, {Vermeulen},
  {Taylor}, {Campbell}, {Pearson}, {Readhead}, {Xu}, {Browne}, {Henstock}, \&
  {Wilkinson}}]{Britzen2007a}
{Britzen}, S., {Vermeulen}, R.~C., {Taylor}, G.~B., {et~al.}
  2007{\natexlab{b}}, \aap, 472, 763

\bibitem[{{Britzen} {et~al.}(1999){Britzen}, {Vermeulen}, {Taylor}, {Pearson},
  {Readhead}, {Wilkinson}, \& {Browne}}]{Britzen1999}
{Britzen}, S., {Vermeulen}, R.~C., {Taylor}, G.~B., {et~al.} 1999, in
  Astronomical Society of the Pacific Conference Series, Vol. 159, BL Lac
  Phenomenon, ed. L.~O. {Takalo} \& A.~{Sillanp{\"a}{\"a}}, 431--+

\bibitem[{{Britzen} {et~al.}(2010{\natexlab{b}}){Britzen}, {Witzel}, {Gong},
  {Zhang}, {Gopal-Krishna}, {Goyal}, {Aller}, {Aller}, \&
  {Zensus}}]{Britzen2010b}
{Britzen}, S., {Witzel}, A., {Gong}, B.~P., {et~al.} 2010{\natexlab{b}}, ArXiv
  e-prints

\bibitem[{{Chang} {et~al.}(2010){Chang}, {Ros}, {Kadler}, {Aller}, {Aller},
  {Angelakis}, {Fuhrmann}, {Nestoras}, \& {Ungerechts}}]{Chang2010}
{Chang}, C.~S., {Ros}, E., {Kadler}, M., {et~al.} 2010, ArXiv e-prints

\bibitem[{{Cohen} {et~al.}(2007){Cohen}, {Lister}, {Homan}, {Kadler},
  {Kellermann}, {Kovalev}, \& {Vermeulen}}]{Cohen2007}
{Cohen}, M.~H., {Lister}, M.~L., {Homan}, D.~C., {et~al.} 2007, \apj, 658, 232

\bibitem[{{de Vaucouleurs}(1991)}]{deVaucouleurs1991}
{de Vaucouleurs}, G. 1991, Science, 254, 1667

\bibitem[{{Fichtel} {et~al.}(1994){Fichtel}, {Bertsch}, {Chiang}, {Dingus},
  {Esposito}, {Fierro}, {Hartman}, {Hunter}, {Kanbach}, {Kniffen}, {Kwok},
  {Lin}, {Mattox}, {Mayer-Hasselwander}, {McDonald}, {Michelson}, {von
  Montigny}, {Nolan}, {Pinkau}, {Radecke}, {Rothermel}, {Sreekumar}, {Sommer},
  {Schneid}, {Thompson}, \& {Willis}}]{Fichtel1994}
{Fichtel}, C.~E., {Bertsch}, D.~L., {Chiang}, J., {et~al.} 1994, \apjs, 94, 551

\bibitem[{{Gehrels} {et~al.}(1993){Gehrels}, {Chipman}, \&
  {Kniffen}}]{Gehrels1993}
{Gehrels}, N., {Chipman}, E., \& {Kniffen}, D.~A. 1993, \aaps, 97, 5

\bibitem[{{Georganopoulos} {et~al.}(2005){Georganopoulos}, {Perlman}, \&
  {Kazanas}}]{Georganopoulos2005}
{Georganopoulos}, M., {Perlman}, E.~S., \& {Kazanas}, D. 2005, \apjl, 634, L33

\bibitem[{{Ghisellini} {et~al.}(1985){Ghisellini}, {Maraschi}, \&
  {Treves}}]{Ghisellini1985}
{Ghisellini}, G., {Maraschi}, L., \& {Treves}, A. 1985, \aap, 146, 204

\bibitem[{{Hartman} {et~al.}(1999){Hartman}, {Bertsch}, {Bloom}, {Chen},
  {Deines-Jones}, {Esposito}, {Fichtel}, {Friedlander}, {Hunter}, {McDonald},
  {Sreekumar}, {Thompson}, {Jones}, {Lin}, {Michelson}, {Nolan}, {Tompkins},
  {Kanbach}, {Mayer-Hasselwander}, {M{\"u}cke}, {Pohl}, {Reimer}, {Kniffen},
  {Schneid}, {von Montigny}, {Mukherjee}, \& {Dingus}}]{Hartman1999}
{Hartman}, R.~C., {Bertsch}, D.~L., {Bloom}, S.~D., {et~al.} 1999, \apjs, 123,
  79

\bibitem[{{Hook} {et~al.}(1995){Hook}, {McMahon}, {Patnaik}, {Browne},
  {Wilkinson}, {Irwin}, \& {Hazard}}]{Hook1995}
{Hook}, I.~M., {McMahon}, R.~G., {Patnaik}, A.~R., {et~al.} 1995, \mnras, 273,
  L63+

\bibitem[{{Hovatta} {et~al.}(2009){Hovatta}, {Valtaoja}, {Tornikoski}, \&
  {L{\"a}hteenm{\"a}ki}}]{Hovatta2009}
{Hovatta}, T., {Valtaoja}, E., {Tornikoski}, M., \& {L{\"a}hteenm{\"a}ki}, A.
  2009, \aap, 494, 527

\bibitem[{{Hoyle}(1966)}]{Hoyle1966}
{Hoyle}, F. 1966, \nat, 209, 751

\bibitem[{{Jorstad} {et~al.}(2001){Jorstad}, {Marscher}, {Mattox}, {Aller},
  {Aller}, {Wehrle}, \& {Bloom}}]{Jorstad2001b}
{Jorstad}, S.~G., {Marscher}, A.~P., {Mattox}, J.~R., {et~al.} 2001, \apj, 556,
  738

\bibitem[{{Kanbach} {et~al.}(1988){Kanbach}, {Bertsch}, {Fichtel}, {Hartman},
  {Hunter}, {Kniffen}, {Hughlock}, {Favale}, {Hofstadter}, \&
  {Hughes}}]{Kanbach1988}
{Kanbach}, G., {Bertsch}, D.~L., {Fichtel}, C.~E., {et~al.} 1988, Space Science
  Reviews, 49, 69

\bibitem[{{Karouzos} {et~al.}(2010){Karouzos}, {Britzen}, {Eckart}, {Witzel},
  \& {Zensus}}]{Karouzos2010}
{Karouzos}, M., {Britzen}, S., {Eckart}, A., {Witzel}, A., \& {Zensus}, A.
  2010, ArXiv e-prints

\bibitem[{{Kellermann} {et~al.}(2004){Kellermann}, {Lister}, {Homan},
  {Vermeulen}, {Cohen}, {Ros}, {Kadler}, {Zensus}, \&
  {Kovalev}}]{Kellermann2004}
{Kellermann}, K.~I., {Lister}, M.~L., {Homan}, D.~C., {et~al.} 2004, \apj, 609,
  539

\bibitem[{{L{\"a}hteenm{\"a}ki} {et~al.}(1999){L{\"a}hteenm{\"a}ki},
  {Valtaoja}, \& {Wiik}}]{Lahteenmaki1999}
{L{\"a}hteenm{\"a}ki}, A., {Valtaoja}, E., \& {Wiik}, K. 1999, \apj, 511, 112

\bibitem[{{Lister} {et~al.}(2009{\natexlab{a}}){Lister}, {Aller}, {Aller},
  {Cohen}, {Homan}, {Kadler}, {Kellermann}, {Kovalev}, {Ros}, {Savolainen},
  {Zensus}, \& {Vermeulen}}]{Lister2009b}
{Lister}, M.~L., {Aller}, H.~D., {Aller}, M.~F., {et~al.} 2009{\natexlab{a}},
  \aj, 137, 3718

\bibitem[{{Lister} {et~al.}(2009{\natexlab{b}}){Lister}, {Homan}, {Kadler},
  {Kellermann}, {Kovalev}, {Ros}, {Savolainen}, \& {Zensus}}]{Lister2009}
{Lister}, M.~L., {Homan}, D.~C., {Kadler}, M., {et~al.} 2009{\natexlab{b}},
  \apjl, 696, L22

\bibitem[{{Liu} {et~al.}(2010){Liu}, {Jiang}, \& {Shen}}]{Liu2010}
{Liu}, Y., {Jiang}, D.~R., \& {Shen}, Z. 2010, ArXiv e-prints

\bibitem[{{Lobanov} \& {Roland}(2005)}]{Lobanov2005}
{Lobanov}, A.~P. \& {Roland}, J. 2005, \aap, 431, 831

\bibitem[{{Lowe} {et~al.}(2007){Lowe}, {Gawro{\'n}ski}, {Wilkinson}, {Kus},
  {Browne}, {Pazderski}, {Feiler}, \& {Kettle}}]{Lowe2007}
{Lowe}, S.~R., {Gawro{\'n}ski}, M.~P., {Wilkinson}, P.~N., {et~al.} 2007, \aap,
  474, 1093

\bibitem[{{Maraschi} {et~al.}(1992){Maraschi}, {Ghisellini}, \&
  {Celotti}}]{Maraschi1992}
{Maraschi}, L., {Ghisellini}, G., \& {Celotti}, A. 1992, \apjl, 397, L5

\bibitem[{{Neshpor} {et~al.}(1998){Neshpor}, {Stepanyan}, {Kalekin}, {Fomin},
  {Chalenko}, \& {Shitov}}]{Neshpor1998}
{Neshpor}, Y.~I., {Stepanyan}, A.~A., {Kalekin}, O.~P., {et~al.} 1998,
  Astronomy Letters, 24, 134

\bibitem[{{Pearson} {et~al.}(1998){Pearson}, {Browne}, {Henstock}, {Polatidis},
  {Readhead}, {Taylor}, {Thakkar}, {Vermeulen}, {Wilkinson}, \&
  {Xu}}]{Pearson1998}
{Pearson}, T.~J., {Browne}, I.~W.~A., {Henstock}, D.~R., {et~al.} 1998, in
  Astronomical Society of the Pacific Conference Series, Vol. 144, IAU Colloq.
  164: Radio Emission from Galactic and Extragalactic Compact Sources, ed.
  J.~A. {Zensus}, G.~B. {Taylor}, \& J.~M. {Wrobel}, 17--+

\bibitem[{{Piner} {et~al.}(2008){Piner}, {Pant}, \& {Edwards}}]{Piner2008}
{Piner}, B.~G., {Pant}, N., \& {Edwards}, P.~G. 2008, \apj, 678, 64

\bibitem[{{Pollack} {et~al.}(2003){Pollack}, {Taylor}, \&
  {Zavala}}]{Pollack2003}
{Pollack}, L.~K., {Taylor}, G.~B., \& {Zavala}, R.~T. 2003, \apj, 589, 733

\bibitem[{{Punch} {et~al.}(1992){Punch}, {Akerlof}, {Cawley}, {Chantell},
  {Fegan}, {Fennell}, {Gaidos}, {Hagan}, {Hillas}, {Jiang}, {Kerrick}, {Lamb},
  {Lawrence}, {Lewis}, {Meyer}, {Mohanty}, {O'Flaherty}, {Reynolds}, {Rovero},
  {Schubnell}, {Sembroski}, {Weekes}, \& {Wilson}}]{Punch1992}
{Punch}, M., {Akerlof}, C.~W., {Cawley}, M.~F., {et~al.} 1992, \nat, 358, 477

\bibitem[{{Pushkarev} {et~al.}(2009){Pushkarev}, {Kovalev}, {Lister}, \&
  {Savolainen}}]{Pushkarev2009}
{Pushkarev}, A.~B., {Kovalev}, Y.~Y., {Lister}, M.~L., \& {Savolainen}, T.
  2009, \aap, 507, L33

\bibitem[{{Roland} {et~al.}(2008){Roland}, {Britzen}, {Kudryavtseva}, {Witzel},
  \& {Karouzos}}]{Roland2008}
{Roland}, J., {Britzen}, S., {Kudryavtseva}, N.~A., {Witzel}, A., \&
  {Karouzos}, M. 2008, \aap, 483, 125

\bibitem[{{Savolainen} {et~al.}(2010){Savolainen}, {Homan}, {Hovatta},
  {Kadler}, {Kovalev}, {Lister}, {Ros}, \& {Zensus}}]{Savolainen2010}
{Savolainen}, T., {Homan}, D.~C., {Hovatta}, T., {et~al.} 2010, \aap, 512, A24+

\bibitem[{{Sikora} {et~al.}(1994){Sikora}, {Begelman}, \& {Rees}}]{Sikora1994}
{Sikora}, M., {Begelman}, M.~C., \& {Rees}, M.~J. 1994, \apj, 421, 153

\bibitem[{{Spergel} {et~al.}(2003){Spergel}, {Verde}, {Peiris}, {Komatsu},
  {Nolta}, {Bennett}, {Halpern}, {Hinshaw}, {Jarosik}, {Kogut}, {Limon},
  {Meyer}, {Page}, {Tucker}, {Weiland}, {Wollack}, \& {Wright}}]{Spergel2003}
{Spergel}, D.~N., {Verde}, L., {Peiris}, H.~V., {et~al.} 2003, \apjs, 148, 175

\bibitem[{{Steffen} {et~al.}(1995{\natexlab{a}}){Steffen}, {Krichbaum},
  {Britzen}, \& {Witzel}}]{Steffen1995b}
{Steffen}, W., {Krichbaum}, T.~P., {Britzen}, S., \& {Witzel}, A.
  1995{\natexlab{a}}, in The XXVIIth Young European Radio Astronomers
  Conference, ed. {D.~A.~Green \& W.~Steffen}, 29--+

\bibitem[{{Steffen} {et~al.}(1995{\natexlab{b}}){Steffen}, {Zensus},
  {Krichbaum}, {Witzel}, \& {Qian}}]{Steffen1995}
{Steffen}, W., {Zensus}, J.~A., {Krichbaum}, T.~P., {Witzel}, A., \& {Qian},
  S.~J. 1995{\natexlab{b}}, \aap, 302, 335

\bibitem[{{Tavani} {et~al.}(2008){Tavani}, {Barbiellini}, {Argan},
  {Bulgarelli}, {Caraveo}, {Chen}, {Cocco}, {Costa}, {de Paris}, {Del Monte},
  {Di Cocco}, {Donnarumma}, {Feroci}, {Fiorini}, {Froysland}, {Fuschino},
  {Galli}, {Gianotti}, {Giuliani}, {Evangelista}, {Labanti}, {Lapshov},
  {Lazzarotto}, {Lipari}, {Longo}, {Marisaldi}, {Mastropietro}, {Mauri},
  {Mereghetti}, {Morelli}, {Morselli}, {Pacciani}, {Pellizzoni}, {Perotti},
  {Picozza}, {Pontoni}, {Porrovecchio}, {Prest}, {Pucella}, {Rapisarda},
  {Rossi}, {Rubini}, {Soffitta}, {Trifoglio}, {Trois}, {Vallazza},
  {Vercellone}, {Zambra}, {Zanello}, {Giommi}, {Antonelli}, \&
  {Pittori}}]{Tavani2008}
{Tavani}, M., {Barbiellini}, G., {Argan}, A., {et~al.} 2008, Nuclear
  Instruments and Methods in Physics Research A, 588, 52

\bibitem[{{Tavecchio} \& {Ghisellini}(2008)}]{Tavecchio2008}
{Tavecchio}, F. \& {Ghisellini}, G. 2008, ArXiv e-prints

\bibitem[{{Taylor} {et~al.}(1996){Taylor}, {Vermeulen}, {Readhead}, {Pearson},
  {Henstock}, \& {Wilkinson}}]{Taylor1996}
{Taylor}, G.~B., {Vermeulen}, R.~C., {Readhead}, A.~C.~S., {et~al.} 1996,
  \apjs, 107, 37

\bibitem[{{Thompson} {et~al.}(1996){Thompson}, {Bertsch}, {Dingus}, {Esposito},
  {Etienne}, {Fichtel}, {Friedlander}, {Hartman}, {Hunter}, {Kendig}, {Mattox},
  {McDonald}, {Mukherjee}, {Ramanamurthy}, {Sreekumar}, {von Montigny},
  {Fierro}, {Jones}, {Lin}, {Michelson}, {Nolan}, {Tompkins}, {Willis},
  {Kanbach}, {Mayer-Hasselwander}, {Merck}, {Pohl}, {Kniffen}, \&
  {Schneid}}]{Thompson1996}
{Thompson}, D.~J., {Bertsch}, D.~L., {Dingus}, B.~L., {et~al.} 1996, \apjs,
  107, 227

\bibitem[{{Vermeulen} {et~al.}(2003){Vermeulen}, {Britzen}, {Taylor},
  {Pearson}, {Readhead}, {Wilkinson}, \& {Browne}}]{Vermeulen2003}
{Vermeulen}, R.~C., {Britzen}, S., {Taylor}, G.~B., {et~al.} 2003, in
  Astronomical Society of the Pacific Conference Series, Vol. 300, Radio
  Astronomy at the Fringe, ed. J.~A. {Zensus}, M.~H. {Cohen}, \& E.~{Ros},
  43--+

\bibitem[{{von Montigny} {et~al.}(1995){von Montigny}, {Bertsch}, {Chiang},
  {Dingus}, {Esposito}, {Fichtel}, {Fierro}, {Hartman}, {Hunter}, {Kanbach},
  {Kniffen}, {Lin}, {Mattox}, {Mayer-Hasselwander}, {Michelson}, {Nolan},
  {Radecke}, {Schneid}, {Sreekumar}, {Thompson}, \& {Willis}}]{Montigny1995}
{von Montigny}, C., {Bertsch}, D.~L., {Chiang}, J., {et~al.} 1995, \apj, 440,
  525

\bibitem[{{Zensus} {et~al.}(1995){Zensus}, {Cohen}, \& {Unwin}}]{Zensus1995}
{Zensus}, J.~A., {Cohen}, M.~H., \& {Unwin}, S.~C. 1995, \apj, 443, 35

\end{thebibliography}

\onecolumn
\appendix
\section{$\gamma$-ray properties of the CJF sample}
\begin{longtable}{l l l l l l l l l l l l}
\caption{CJF sources detected in the $\gamma$-ray regime by Fermi-LAT.}
\label{tab:gamma}
\\
\hline\\
\textbf{Source}	&	\textbf{Type}	&\textbf{z}	&$\mathbf{F_{\gamma}}$			 &$\mathbf{\delta F_{\gamma}}$ &$\alpha_{\gamma}$ &$\delta\alpha_{\gamma}$ &$\log{(\nu L_{\nu})_{\gamma}}$ & $\delta\log{(\nu L_{\nu})_{\gamma}}$	 &\textbf{Var} &$\beta_{app}^{max}$ &$\delta\beta_{app}^{max}$	\\
       &			   	    &		        &\multicolumn{2}{c}{($10^{-9}\:\mathrm{ph}\:\mathrm{cm}^{-2}\:\mathrm{s}^{-1})$} & & &\multicolumn{2}{c}{($erg$)}	& 	& &	    \\
\hline\\
\endfirsthead

\multicolumn{12}{l}{{\tablename} \thetable{} -- Continued} \\
\hline\\
\textbf{Source}	&	\textbf{Type}	&\textbf{z}	&$\mathbf{F_{\gamma}}$			 &$\mathbf{\delta F_{\gamma}}$ &$\alpha_{\gamma}$ &$\delta\alpha_{\gamma}$ &$\log{(\nu L_{\nu})_{\gamma}}$ & $\delta\log{(\nu L_{\nu})_{\gamma}}$	 &\textbf{Var} &$\beta_{app}^{max}$ &$\delta\beta_{app}^{max}$	\\
       &			   	    &		        &\multicolumn{2}{c}{($10^{-9}\:\mathrm{ph}\:\mathrm{cm}^{-2}\:\mathrm{s}^{-1})$} & & &\multicolumn{2}{c}{($erg\:\mathrm{s}^{-1}$)}	& 	& &	    \\
\hline\\
\endhead

\hline\\
\multicolumn{12}{l}{{Continued on Next Page\ldots}} \\
\endfoot

\\
\endlastfoot
0003+380	    &G(Q)	&0.229	 &0.6	   &0.3	   & 2.86	&0.13	 &44.57	    &0.22    &N	      &4.899	&0     \\
0110+495        &Q	&0.389   &0.7	   &0.3    & 2.29	&0.18    &45.33	    &0.19    &N       &1.049	&0.646 \\
0133+476        &Q	&0.859   &9.6	   &0.6    & 2.34	&0.03    &47.33	    &0.03    &Y       &10.986	&2.817 \\
0212+735        &Q	&2.367   &1	   &0.4    & 2.85	&0.13    &47.63	    &0.17    &Y       &16.068	&7.121 \\
0218+357        &G(Q)	&0.936   &6.4	   &0.5    & 2.33	&0.04    &47.25	    &0.03    &Y       &n/a	&n/a	\\
0219+428        &BL	&0.444   &24.9	   &1      & 1.93	&0.02    &47.15	    &0.02    &Y       &14.04	&5.03  \\
0227+403        &Q	&1.019   &1.4	   &0.3    & 2.43	&0.13    &46.68	    &0.09    &Y       &4.013	&1.07  \\
0307+380        &Q	&0.816   &0.6	   &0.3    & 2.49	&0.15    &46.05	    &0.22    &Y       &4.283	&0     \\
0316+413        &G	&0.018   &17.3	   &0.8    & 2.13	&0.02    &43.89	    &0.02    &Y       &0.978	&0.204 \\
0346+800        &Q(?)	&n/a     &2	   &0.3    & 2.5	&0.08    &n/a	    &n/a     &Y       &n/a	&n/a	\\
0621+446        &BL	&n/a     &1	   &0.3    & 2.03	&0.18    &n/a	    &n/a     &N       &n/a	&n/a	\\
0633+734        &Q	&1.85    &0.6	   &0.3    & 2.73	&0.17    &47.06	    &0.22    &N       &21.673	&1.036 \\
0650+453        &Q	&0.933   &6.1	   &0.5    & 2.32	&0.04    &47.23	    &0.04    &Y       &n/a	&n/a	\\
0707+476        &Q(BL)	&1.292   &0.9	   &0.3    & 2.51	&0.13    &46.77	    &0.14    &N       &4.248	&4.439 \\
0716+714        &BL	&0.31    &13.1	   &0.7    & 2.15	&0.03    &46.38	    &0.02    &Y       &n/a	&n/a	\\
0749+540        &BL	&n/a     &1.2	   &0.3    & 1.95	&0.16    &n/a	    &n/a     &N       &n/a	&n/a	\\
0800+618        &Q	&3.044   &0.6	   &0.2    & 2.83	&0.13    &47.74	    &0.14    &Y       &10.003	&9.591 \\
0814+425        &BL	&0.53    &8.7	   &0.6	   & 2.15	&0.04	 &46.01	    &0.03    &Y		&2.72	&1.834 \\
0820+560        &Q	&1.409   &0.9	   &0.3    & 2.87	&0.11    &46.91	    &0.14    &Y       &5.311	&0     \\
0836+710        &Q	&2.18    &1.2	   &0.3    & 2.98	&0.12    &47.64	    &0.11    &N       &30.59	&4.457 \\
0917+449        &Q	&2.18    &14	   &0.7    & 2.28	&0.02    &48.58	    &0.02    &Y       &14.246	&5.506 \\
0917+624        &Q	&1.446   &1.1	   &0.3    & 2.7	&0.15    &47.01	    &0.12    &N       &3.214	&1.026 \\
0925+504        &BL	&0.37    &0.7	   &0.2    & 1.91	&0.23    &45.43	    &0.12    &N       &n/a	&n/a	\\
0954+556        &Q	&0.895   &10.5	   &0.6    & 2.05	&0.03    &47.50	    &0.02    &N       &n/a	&n/a	\\
0954+658        &BL	&0.368   &0.5	   &0.3    & 2.51	&0.16    &45.06	    &0.26    &N       &9.874	&0     \\
1015+359        &Q	&1.226   &0.5	   &0.2    & 2.71	&0.15    &46.46	    &0.17    &N       &10.273	&1.59  \\
1020+400        &Q	&1.254   &0.6	   &0.2    & 2.45	&0.17    &46.56	    &0.14    &N       &13.169	&5.094 \\
1030+415        &Q	&1.12    &1.1	   &0.3    & 2.48	&0.12    &46.69	    &0.12    &N       &4.186	&2.408 \\
1030+611        &Q	&1.401   &2.2	   &0.3    & 2.46	&0.08    &45.74	    &0.06    &Y       &1.879	&1.817 \\
1039+811        &Q	&1.254   &0.9	   &0.2    & 2.95	&0.13    &46.77	    &0.10    &Y       &10.125	&1.553 \\
1044+719        &Q	&1.15    &1.6	   &0.3    & 2.47	&0.13    &46.88	    &0.08    &N       &3.039	&0.643 \\
1101+384        &BL	&0.031   &26.1	   &1      & 1.81	&0.02    &44.73	    &0.02    &Y       &0.187	&0.066 \\
1144+402        &Q	&1.088   &1	   &0.3    & 2.47	&0.13    &46.61	    &0.13    &N       &n/a	&n/a	\\
1206+415        &BL	&n/a     &0.5	   &0.2    & 1.85	&0.22    &n/a	    &n/a     &N       &n/a	&n/a	\\
1221+809        &BL	&n/a     &1	   &0.3    & 2.27	&0.14    &n/a	    &n/a     &N       &n/a	&n/a	\\
1246+586        &BL	&n/a     &4.5	   &0.4    & 2.18	&0.06    &n/a	    &n/a     &N       &n/a	&n/a	\\
1250+532        &BL	&n/a     &3	   &0.4    & 2.14	&0.07    &n/a	    &n/a     &N       &n/a	&n/a	\\
1306+360        &Q	&1.055   &2.2	   &0.3    & 2.3	&0.09    &46.93	    &0.06    &Y       &4.611	&0     \\
1322+835        &na	&1.024   &0.5	   &0.2    & 2.49	&0.22    &46.24	    &0.17    &N       &n/a	&n/a	\\
1357+769        &BL(Q)	&n/a     &1.1	   &0.2    & 2.25	&0.16    &n/a	    &n/a     &N       &n/a	&n/a	\\
1418+546        &BL	&0.151   &0.9	   &0.2    & 2.77	&0.17    &44.34	    &0.10    &N       &2.562	&0     \\
1432+422        &Q	&1.24    &0.7	   &0.2    & 2.25	&0.2     &46.63	    &0.12    &N       &17.69	&0.678 \\
1504+377        &G(Q)	&0.672   &0.7	   &0.2    & 2.59	&0.13    &45.87	    &0.12    &Y       &10.547	&1.655 \\
1633+382        &Q	&1.807   &6.8	   &0.5    & 2.47	&0.04    &48.05	    &0.03    &Y       &12.18	&4.951 \\
1641+399        &Q	&0.595   &5.6	   &0.5    & 2.49	&0.04    &46.65	    &0.04    &Y       &6.114	&0     \\
1652+398        &BL	&0.034   &8.3	   &0.6    & 1.85	&0.04    &44.29	    &0.03    &Y       &0.898	&0     \\
1700+685        &G(Q)	&0.301   &3.7	   &0.4    & 2.28	&0.06    &45.78	    &0.05    &Y       &1.037	&1.131 \\
1722+401        &Q(G)	&1.049   &2.9	   &0.4    & 2.37	&0.07    &47.04	    &0.06    &Y       &20.883	&10.441\\
1726+455        &Q	&0.717   &1.2	   &0.3    & 2.57	&0.09    &46.19	    &0.11    &Y       &5.686	&0.812 \\
1732+389        &Q	&0.97    &6	   &0.5    & 2.22	&0.05    &47.29	    &0.04    &Y       &12.065	&1.805 \\
1739+522        &Q	&1.381   &3.4	   &0.4    & 2.71	&0.05    &47.44	    &0.05    &Y       &11.475	&9.021 \\
1749+701        &BL	&0.77    &2	   &0.3    & 2.05	&0.1     &46.60	    &0.07    &N       &14.805	&4.992 \\
1747+433        &BL	&n/a     &2.3	   &0.3    & 2.12	&0.09    &n/a	    &n/a     &N       &n/a	&n/a	\\
1803+784        &BL	&0.68    &3	   &0.4    & 2.35	&0.07    &46.56	    &0.06    &Y       &13.567	&2.993 \\
1807+698        &BL	&0.051   &1.9	   &0.3    & 2.6	&0.08    &43.67	    &0.07    &N       &0.398	&0     \\
1823+568        &BL	&0.664   &2.7	   &0.4    & 2.34	&0.07    &46.49	    &0.06    &Y       &8.994	&1.524 \\
1849+670        &Q	&0.657   &13.3	   &0.7    & 2.25	&0.03    &47.19	    &0.02    &Y       &8.65	&0.567 \\
1851+488        &Q	&1.25    &0.9	   &0.3    & 2.6	&0.12    &46.73	    &0.14    &N       &n/a	&n/a	\\
2007+777        &BL	&0.342   &1.4	   &0.3    & 2.42	&0.16    &45.45	    &0.09    &N       &1.953	&0.849 \\
2010+723        &BL	&n/a     &1.6	   &0.4    & 2.45	&0.15    &n/a	    &n/a     &N       &n/a	&n/a	\\
2023+760        &BL	&n/a     &1.1	   &0.3    & 2.52	&0.18    &n/a	    &n/a     &N       &n/a	&n/a	\\
2200+420        &BL	&0.069   &7.1	   &0.6    & 2.38	&0.04    &44.59	    &0.04    &Y       &3.15	&0.305 \\
\hline
\multicolumn{12}{l}{\tablefoot{ Columns (1)-(3) give the IAU name, type, and redshift of the source. In brackets is given the alternative classification from the Fermi-LAT team, where applicable. Column(4) gives the $\gamma$-ray flux (detections are from the first Fermi-LAT catalog, \citealt{Abdo2010b}), Col. (5) gives the uncertainty for the $\gamma$-ray fluxes, Cols. (6) and (7) give the $\gamma$ photon index and its uncertainty, Cols. (8) and (9) give the calculated $\gamma$-ray luminosity (in the $\log{\nu L_{\nu}}$ form and in logarithmic scale) and the respective uncertainty, Col.(10) denotes sources with detected variability in $\gamma$-rays, and Cols. (11) and (12) show the maximum apparent jet speed for that source and its uncertainty.}}
\end{longtable}
\twocolumn




\end{document}